%% file: TTL_master.tex
\documentclass[reprint,aps,prd,longbibliography,floatfix,superscriptaddress,showkeys]{revtex4-2} 

\usepackage{amsmath}
\usepackage{amsfonts}
\usepackage{amssymb}

\usepackage{url}
\usepackage{hyperref}

\usepackage[dvipsnames]{xcolor}

\definecolor{tolblue}{RGB}{68,119,170}

\usepackage{tikz}

\usepackage{booktabs} %

\usepackage[normalem]{ulem} %

\begin{document}

\title{Post-processing subtraction of tilt-to-length noise in LISA in the presence of gravitational wave signals}

\def\addressmpg{Max Planck Institute for Gravitational Physics (Albert-Einstein-Institut), 
30167 Hannover, Germany}
\def\addressluh{Leibniz Universit\"at Hannover, 30167 Hannover, Germany}
\def\addressuf{Department of Physics, 2001 Museum Road, University of Florida, Gainesville, Florida
32611, USA}

\author{M-S~Hartig}\email{marie-sophie.hartig@aei.mpg.de}\affiliation{\addressluh}\affiliation{\addressmpg}
\author{S~Paczkowski}\email{sarah.paczkowski@aei.mpg.de}\affiliation{\addressmpg}\affiliation{\addressluh}
\author{M~Hewitson}\thanks{Now at the European Space Technology Centre, European Space Agency, 2200 AG Noordwijk, The Netherlands}\affiliation{\addressmpg}\affiliation{\addressluh}
\author{G~Heinzel}\affiliation{\addressmpg}\affiliation{\addressluh}
\author{G~Wanner}\affiliation{\addressluh}\affiliation{\addressmpg}

\date{\today}

\begin{abstract}
The Laser Interferometer Space Antenna (LISA) will be the first space-based gravitational wave (GW) observatory. It will measure gravitational wave signals in the frequency regime from 0.1\,mHz to 1\,Hz. %
The success of these measurements will depend on the suppression of the various instrument noises.
One important noise source in LISA will be tilt-to-length (TTL) coupling. Here, it is understood as the coupling of angular jitter, predominantly from the spacecraft, into the interferometric length readout. 
The current plan is to subtract this noise in-flight in post-processing as part of a noise minimization strategy.
It is crucial to distinguish TTL coupling well from the GW signals in the same readout to ensure that the noise will be properly modeled. 
Furthermore, it is important that the subtraction of TTL noise will not degrade the GW signals.
In the present manuscript, we show on simulated LISA data and for four different GW signal types that the GW responses have little effect on the quality of the TTL coupling fit and subtraction.
Also, the GW signal characteristics were not altered by the TTL coupling subtraction.
\end{abstract}

\maketitle

\input{content/introduction.tex}

\input{content/TTL_LISA.tex}

\input{content/TTL_Simulation.tex}

\input{content/TTL_Orbits.tex}

\input{content/TTL_GW.tex}

\input{content/summary.tex}

\input{content/acknowledgements.tex}

\bibliographystyle{unsrt}
\bibliography{References.bib}

\appendix
\input{content/appendix.tex}

\end{document}

%% file: content/introduction.tex
\section{Introduction}

The ESA-led Laser Interferometer Space Antenna (LISA) mission, scheduled for launch in the next decade, will be a space-based gravitational wave (GW) observatory \cite{Danzmann2011,Redbook2024}.
The scope of the LISA mission is to measure GW signals in the frequency regime from 0.1\,mHz to 1\,Hz. The primary targeted sources are galactic binaries, verification binaries, stellar mass and massive black hole binaries, an extreme mass ratio inspiral and potential multi-band sources \cite{Redbook2024}.
LISA will use laser interferometry for the signal detection. The GW signals will be extracted from the interferometric measurements of the distance changes between three pairs of test masses (TMs), each pair hosted by a spacecraft (SC). The three SC will be 2.5~million kilometers apart, each in a heliocentric orbit, forming an approximately equilateral triangle.
One of the major noise sources in the interferometric distance measurement will be the coupling of angular spacecraft (SC) and movable optical sub-assembly (MOSA) jitter. 
We refer to this noise as tilt-to-length (TTL) coupling \cite{Hartig22,Hartig23}.
In this work, we will show that the TTL noise can be well fitted and subtracted from the data in the presence of GW signals.

In the first instance, TTL coupling will be directly reduced by design and realignment. To suppress the remaining TTL noise below the other instrument noises, it will be required to fit and subtract it in post-processing.
The subtraction strategy has been successfully tested during the LISA Pathfinder mission \cite{Wanner2017,Armano2023_TTL} and for simulated LISA data \cite{Paczkowski2022,Paczkowski2024}.
The authors of \cite{George2022,Hartig2024} have studied the precision of the coupling coefficient estimation for daily data.
In addition, maneuver designs for a better coupling coefficient estimation have been investigated \cite{Houba2022a,Houba2022b,Wegener2024}.
However, only some of these analyses have considered GW signals.
\cite{George2022,Hartig2024} included different GW sources in their precision analyses and \cite{Houba2022b} incorporated them in their maneuver design studies.
In our paper we verify the suppression of the TTL coupling by subtraction.

This paper is structured as follows:
In section~\ref{sec:TTL_LISA}, we will describe the TTL coupling in the LISA mission and introduce the notations used in this manuscript.
In section~\ref{sec:Simulation}, we present the simulation environment and the analysis scheme used.
In preparation for the main investigations of this work, we perform preliminary tests with changing arm lengths, see section~\ref{sec:Orbits}.
Then, we evaluate in section~\ref{sec:GW} how well the TTL coupling coefficient can estimated and analyze the TTL noise minimization by subtraction in the presence of different GW signal types. 
Finally, we summarize our results in section~\ref{sec:summary}.

%% file: content/TTL_LISA.tex
\section{Tilt-to-length noise in LISA}
\label{sec:TTL_LISA}

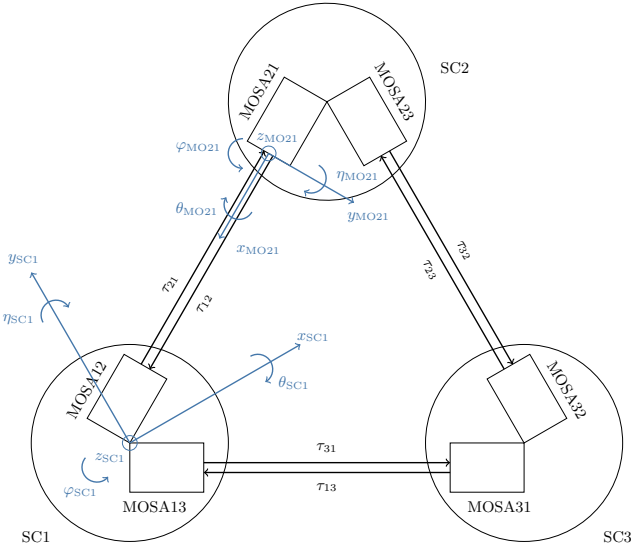
\begin{figure}
\resizebox{\columnwidth}{!}{
\begin{tikzpicture}
\draw (0,0) circle (2);
\draw (-1.9,-1.9) node {SC1};
\draw[rotate around={60:(0,0)}] (8,0) circle (2);
\draw[rotate around={60:(0,0)}] (9.9,-1.9) node {SC2};
\draw (8,0) circle (2);
\draw (9.9,-1.9) node {SC3};
\draw[rotate around={60:(0,0)}] (0,0) rectangle (1.5,1);
\draw[rotate around={60:(0,0)}] (0.5,1.3) node[rotate around={60:(0,0)}] {MOSA12}; 
\draw (0,-1) rectangle (1.5,0);
\draw (0.5,-1.3) node {MOSA13};
\draw[rotate around={60:(0,0)}] (6.5,0) rectangle (8,1);
\draw[rotate around={60:(0,0)}] (7.5,1.3) node[rotate around={60:(0,0)}] {MOSA21}; 
\draw[rotate around={-60:(8,0)}] (0,0) rectangle (1.5,1);
\draw[rotate around={-60:(8,0)}] (0.5,1.3) node[rotate around={-60:(0,0)}] {MOSA23}; 
\draw[rotate around={30:(8,0)}] (8,0) rectangle (9,1.5);
\draw[rotate around={-60:(8,0)}] (7.5,1.3) node[rotate around={-60:(0,0)}] {MOSA32}; 
\draw (6.5,-1) rectangle (8,0);
\draw (7.5,-1.3) node {MOSA31};
\draw[thick, ->,rotate around={60:(0,0)}] (1.5,0.6) -- (6.5,0.6);
\draw[rotate around={60:(0,0)}] (3.2,0.9) node[rotate around={60:(0,0)}] {$\tau_{21}$};
\draw[thick, ->,rotate around={60:(0,0)}] (6.5,0.4) -- (1.5,0.4);
\draw[rotate around={60:(0,0)}] (3.2,0.1) node[rotate around={60:(0,0)}] {$\tau_{12}$};
\draw[thick, ->] (1.5,-0.4) -- (6.5,-0.4);
\draw (4,-0.1) node {$\tau_{31}$};
\draw[thick, ->] (6.5,-0.6) -- (1.5,-0.6);
\draw (4,-0.9) node {$\tau_{13}$};
\draw[thick, ->,rotate around={-60:(8,0)}] (6.5,0.4) -- (1.5,0.4);
\draw[rotate around={-60:(8,0)}] (4,0.1) node[rotate around={-60:(0,0)}] {$\tau_{23}$};
\draw[thick, ->,rotate around={-60:(8,0)}] (1.5,0.6) -- (6.5,0.6);
\draw[rotate around={-60:(8,0)}] (4,0.9) node[rotate around={-60:(0,0)}] {$\tau_{32}$};
\draw[tolblue, thick, ->,rotate around={30:(0,0)}] (0,0) -- (4,0); 
\draw[tolblue,rotate around={30:(0,0)}] (4.3,0) node {$x_{\mathrm{SC}1}$};
\draw[tolblue, thick, ->,rotate around={30:(0,0)}] (0,0) -- (0,4); 
\draw[tolblue,rotate around={30:(0,0)}] (0,4.3) node {$y_{\mathrm{SC}1}$};
\draw[tolblue] (0,0) circle (0.15);
\draw[tolblue] (-0.4,-0.3) node {$z_{\mathrm{SC}1}$};
\draw[tolblue, thick, ->,rotate around={30:(0,0)}] (3,0.3) arc (90:-90:0.3);
\draw[tolblue,rotate around={30:(0,0)}] (3.5,-0.6) node {$\theta_\mathrm{SC1}$};
\draw[tolblue, thick, ->,rotate around={30:(0,0)}] (-0.3,3) arc (180:0:0.3);
\draw[tolblue,rotate around={30:(0,0)}] (-0.7,3.3) node {$\eta_\mathrm{SC1}$};
\draw[tolblue, thick, ->] (-0.9,-0.3) arc (140:320:0.3);
\draw[tolblue] (-1,-1) node {$\varphi_\mathrm{SC1}$};
\draw[tolblue, thick, ->,rotate around={60:(0,0)}] (6.5,0.5) -- (4.5,0.5);
\draw[tolblue, rotate around={60:(0,0)}] (4.7,-0.3) node {$x_{\mathrm{MO}21}$};
\draw[tolblue, thick, ->,rotate around={60:(0,0)}] (6.5,0.5) -- (6.5,-1.5);
\draw[tolblue, rotate around={60:(0,0)}] (6.4,-1.9) node {$y_{\mathrm{MO}21}$};
\draw[tolblue, rotate around={60:(0,0)}] (6.5,0.5) circle (0.15);
\draw[tolblue] (3.,6.2) node {$z_{\mathrm{MO}21}$};
\draw[tolblue, thick, ->,rotate around={60:(0,0)}] (5.3,0.2) arc (270:90:0.3);
\draw[tolblue,rotate around={60:(0,0)}] (4.8,1.2) node {$\theta_\mathrm{MO21}$};
\draw[tolblue, thick, ->,rotate around={60:(0,0)}] (6.8,-0.5) arc (360:180:0.3);
\draw[tolblue,rotate around={60:(0,0)}] (7.0,-1.3) node {$\eta_\mathrm{MO21}$};
\draw[tolblue, thick, ->,rotate around={60:(0,0)}] (6.5,1.1) arc (30:210:0.3);
\draw[tolblue,rotate around={60:(0,0)}] (5.9,1.8) node {$\varphi_\mathrm{MO21}$};
\end{tikzpicture} }
\caption{Naming conventions for the LISA SC and MOSAs. Each SC and MOSA has its own coordinate system. We show them exemplary for SC1 and MOSA21. The $\tau_{ij},\ i,j\in\{1,2,3\}$ are the light propagation times between the SC. For simplicity, the TMs are not shown in this figure. The figure has been adapted from \cite{Hartig2024}.}
\label{fig:LISA_notation}
\end{figure}

In figure~\ref{fig:LISA_notation}, we show a sketch of the LISA constellation. All three SC send laser beams through telescopes to both other SC. The beams will be received by the same telescopes at the distant SC. Then, they will propagate through an optical setup and interfere with a local beam.
The telescope and the optical bench together form the moving optical sub-assembly (MOSA), of which we have a total of six on the three SC, see figure~\ref{fig:LISA_notation}. 
In addition to this measurement of the distance changes between the SC, the distance changes between three pairs of TMs and their host SC will be measured (not shown in the figure).

The SC and the MOSAs will be subject to unwanted angular jitter. The jitter will couple into the interferometric length readout and this is what we call TTL coupling.
We will describe it mathematically below.
Since the jitter magnitudes are expected to be in the order of nanoradians, a linear coupling model is considered to be sufficient to describe the non-negligible TTL coupling.

\subsection{Tilt-to-length coupling for a single link} 

In LISA, laser interferometry will be used to measure the distance changes $x_{ij},\ i\neq j\in\{1,2,3\}$, between two TMs on two different SC. A GW signal $h_{ij}$ will decrease and increase the light propagation time between the two SC and can thus be measured.
However, the measurement will also be disturbed by TTL coupling $x_{\mathrm{TTL}_{ij}}$ and other instrument noise sources $n_{ij}$.
All together, we have
\begin{eqnarray}
  x_{ij} = h_{ij} + x_{\mathrm{TTL}_{ij}} + n_{ij} \,.
\end{eqnarray}
For the subtraction of TTL coupling, we fit a linear model to these length measurements using the angle measurements via differential wavefront sensing (DWS) \cite{Morrison1994,Wanner2012}
\begin{eqnarray}
  \alpha^\mathrm{DWS}_{ij} = \alpha_{ij} + n^\mathrm{DWS} , \quad \alpha\in\{\eta,\varphi\}\,.
\label{eq:DWS}
\end{eqnarray}
For a single LISA link, the TTL coupling in SC $i$ pointing to SC $j$ can be written as follows \cite{Paczkowski2022} -- note that the jitter of the local and the remote SC will couple into the length readout:
\begin{eqnarray}
	\hat{x}_{\mathrm{TTL}_{ij}} &=& 
	C_{ij \varphi \rm Rx} \, \varphi_{ij}^\mathrm{DWS} +
	C_{ij \eta \rm Rx} \, \eta_{ij}^\mathrm{DWS} \nonumber\\
	 &&+
	C_{ji \varphi \rm Tx} \, \mathcal{D}_{ij} \varphi_{ji}^\mathrm{DWS} +
	C_{ji \eta \rm Tx} \, \mathcal{D}_{ij} \eta_{ji}^\mathrm{DWS} \,.
\label{eq:TTL_singlelink}
\end{eqnarray}
Here, the $\hat{\ }$ marks that this is an estimate of the true TTL coupling $x_{\mathrm{TTL}_{ij}}$.
The $C$'s denote the TTL coupling coefficients (units: m/rad), which scale the jitter coupling into the length measurement. 
The angles $\varphi_{ij}^\mathrm{DWS}$ and $\eta_{ij}^\mathrm{DWS}$ describe the measured sum of the jitter of the `receiving' SC~$i$ and the MOSA~$ij$, compare figure~\ref{fig:LISA_notation}.
The jitter of the `transmitting' SC~$j$ and MOSA~$ji$ are delayed by the light traveling time $\tau_{ij}$, which is described mathematically by the delay operator $\mathcal{D}_{ij}$.
In total, there are six links between the SC, $ij\in\{12,23,31,21,32,13\}$. This yields in total 24 TTL coupling terms.

\subsection{Time delay interferometry combinations}

In LISA, laser frequency noise will be the dominating noise source in the single link measurement. It will be several orders of magnitude above the LISA requirement, making it impossible to study GW signals in a single link. 
To suppress this noise, a strategy called time-delay interferometry (TDI) \cite{Tinto2021,Tinto2023} will be applied. 
It describes the combination of delayed single link measurements to form a virtual equal arm interferometer.
Most common is the 2nd generation Michelson TDI combination \cite{Tinto2021}. It suppresses the laser frequency noise for varying arm lengths.
The TTL noise in the single link measurements will be traced through TDI as well. 
We find for TDI\,X
\begin{widetext}
\begin{eqnarray}
	\mathrm{TDIX}(\hat{x}_{\mathrm{TTL}_{ij}}) &=&  
	\left(1 -\mathcal{D}_{12}\mathcal{D}_{21} -\mathcal{D}_{12}\mathcal{D}_{21}\mathcal{D}_{13}\mathcal{D}_{31} +\mathcal{D}_{13}\mathcal{D}_{31}\mathcal{D}_{12}\mathcal{D}_{21}\mathcal{D}_{12}\mathcal{D}_{21}\right) \nonumber\\
	&&\cdot \left( C_{13\eta\rm Rx}\,\eta_{13} + C_{31\eta\rm Tx}\,\mathcal{D}_{13}\eta_{31} + C_{31\eta\rm Rx}\,\mathcal{D}_{13}\eta_{31} + C_{13\eta\rm Tx}\,\mathcal{D}_{131}\eta_{13} \right. \nonumber\\
	&&\ \ \left. +C_{13\varphi\rm Rx}\,\varphi_{13} + C_{31\varphi\rm Tx}\,\mathcal{D}_{13}\varphi_{31} + C_{31\varphi\rm Rx}\,\mathcal{D}_{13}\varphi_{31} + C_{13\varphi\rm Tx}\,\mathcal{D}_{131}\varphi_{13} \right) \nonumber\\
	&-& \left(1 -\mathcal{D}_{13}\mathcal{D}_{31} -\mathcal{D}_{13}\mathcal{D}_{31}\mathcal{D}_{12}\mathcal{D}_{21} +\mathcal{D}_{12}\mathcal{D}_{21}\mathcal{D}_{13}\mathcal{D}_{31}\mathcal{D}_{13}\mathcal{D}_{31}\right) \nonumber\\
	&&\cdot \left( C_{12\eta\rm Rx}\,\eta_{12} + C_{21\eta\rm Tx}\,\mathcal{D}_{12}\eta_{21} + C_{21\eta\rm Rx}\,\mathcal{D}_{12}\eta_{21} + C_{12\eta\rm Tx}\,\mathcal{D}_{121}\eta_{12}  \right. \nonumber\\
	&&\ \ \left. +C_{12\varphi\rm Rx}\,\varphi_{12} + C_{21\varphi\rm Tx}\,\mathcal{D}_{12}\varphi_{21} + C_{21\varphi\rm Rx}\,\mathcal{D}_{12}\varphi_{21} + C_{12\varphi\rm Tx}\,\mathcal{D}_{121}\varphi_{12} \right) \,.
\label{eq:TTL_TDI}
\end{eqnarray}
\end{widetext}
The TDI\,Y and TDI\,Z combinations can be found by permutation of the indices. 
A detailed modeling of TTL coupling in the single links and the TDI Michelson observables is presented in \cite{Wanner2024}.
In this work, we will use the TDI\,AET variables \cite{Prince2002} in the computation of the TTL coupling coefficients. TDI\,AET are computed from the 2nd generation TDI\,XYZ variables via
\begin{eqnarray}
  A &=& \frac{1}{\sqrt{2}}\left(Z-X\right) \,, \\
  E &=& \frac{1}{\sqrt{6}}\left(X-2\,Y+Z\right) \,, \\
  T &=& \frac{1}{\sqrt{3}}\left(X+Y+Z\right) \,.
\end{eqnarray}
While the noises are highly correlated in TDI\,XYZ, the TDI\,AET variables are orthogonal and uncorrelated in the idealized case of an equal arm interferometer with equal noise levels in each arm \cite{Baghi2023}.

%% file: content/TTL_Simulation.tex
\section{Simulator and simulation settings}
\label{sec:Simulation}

For the fit and subtraction of the TTL coupling, we will use the scheme first introduced in \cite{Paczkowski2022} and further improved in \cite{Paczkowski2024}.
We consider simulated data for one day.
First, we compute the Fourier transformations of the TDI\,AET variables of the interferometric length measurements including the TTL coupling noise. 
The same is done for the TDI\,AET variables of the angular couplings into the length measurements, i.e.\ the 12 measurements $\alpha_{ij},\ \alpha\in\{\eta,\varphi\},\ i\neq j\in\{1,2,3\}$ and their 12 delays along the corresponding arm. For the derivation of the angular coupling terms, we effectively propagate equation~\eqref{eq:TTL_singlelink} through TDI, setting all coupling coefficients but one to zero.
An MCMC algorithm is then used to fit the coupling coefficients to these two data sets.
Like in \cite{Paczkowski2022,Paczkowski2024}, we chose the TTL coefficients to be 2.3\,mm/rad in all test cases. 
All fits were applied in the frequency range from 3\,mHz to 0.9\,Hz. 
We whitened the data iteratively. Therefore, we considered the LISA displacement noise and TM acceleration noise, but no prior knowledge of the GW signals included in the simulated data. 
Further details on the fitting algorithm and the data preparation are given in \cite{Paczkowski2024}.

In \cite{Paczkowski2022}, the LISASim simulator \cite{LISASim} was used to generate the interferometer data. 
This simulator does not yet include GW signals that we want to study here. 
Therefore, we used the LISANode simulator \cite{LISANode,Bayle2023}, which computes the response of given GW signals in dependence on the orbital position of the SC.
In LISANode, the length differences are given as frequency fluctuations, unlike LISASim, where they are given as length variations.
To use the same code as in \cite{Paczkowski2022}, we converted the frequency data into length data.
Even though LISANode simulates anti-alias filters, we applied an additional low-pass filter to enhance the robustness. In this step, we downsampled the data from 4\,Hz to 2\,Hz
The same jitter levels and shapes have been implemented in both simulators. Moreover, only minor differences exist for the instrument noise contributors, such that we find an overall comparable level of instrument noise in the data from both simulators.
The jitter and noise shapes used in our work are summarized in appendix~\ref{app:noise_and_jitter}.
We also find no significant differences in the accuracy of the TTL coupling coefficient estimation and the level of residual noise after TTL coupling subtraction.  
A more detailed comparison of the results using either of the two simulators is given in \cite{Paczkowski2024}.

An optional difference between the two simulators is that LISANode allows the use of orbit files. I.e.\ the distances between the SC are computed from this orbit information, while only static arm lengths are implemented in LISASim.
We will show the performance of TTL coupling subtraction for LISANode data and varying arm lengths in the next section.

We preprocessed the LISANode data with the python package PyTDI \cite{pyTDI}.
In the next step, we load this preprocessed data in MATLAB and proceed with the TTL noise fit as introduced in \cite{Paczkowski2022,Paczkowski2024}.

%% file: content/TTL_Orbits.tex
\section{TTL coupling estimation for changing arm lengths}
\label{sec:Orbits}

In \cite{Paczkowski2022} a fixed, non-rotating constellation was used. 
This assumption was considered valid since the arm length will change by a maximum of 0.035\% in a day \cite{Paczkowski2022}. %
We will confirm this assumption below by showing that the coefficient estimation and TTL noise subtraction works equally well assuming static or time-variant arms.

For our study case, we generated three LISANode data sets with 
\begin{enumerate}
  \item static arms with beam propagation times: $\tau_{12}\approx8.425\,\rm s$, $\tau_{13}\approx8.322\,\rm s$, $\tau_{23}\approx8.372\,\rm s$, $\tau_{21}\approx8.322\,\rm s$, $\tau_{31}\approx8.322\,\rm s$, $\tau_{32}\approx8.372\,\rm s$;
  \item changing arm lengths gained from Keplerian orbit files \cite{LISAorbits};
  \item changing arm lengths gained from ESA LISA science orbit files \cite{LISAorbits,ESAorbits}.
\end{enumerate}
The static arm lengths given are those assumed in \cite{Paczkowski2022}. While the lengths for the individual links are within the LISA length variations, the variations for the full constellation were chosen for study purposes and may not be particularly realistic.
We find larger arm length variation in the ESA LISA science orbits than for the Keplerian orbits.
All instrument parameters were set the same in all three cases and we assumed identical noise and jitter realizations.

\begin{figure}\includegraphics[width=\columnwidth]{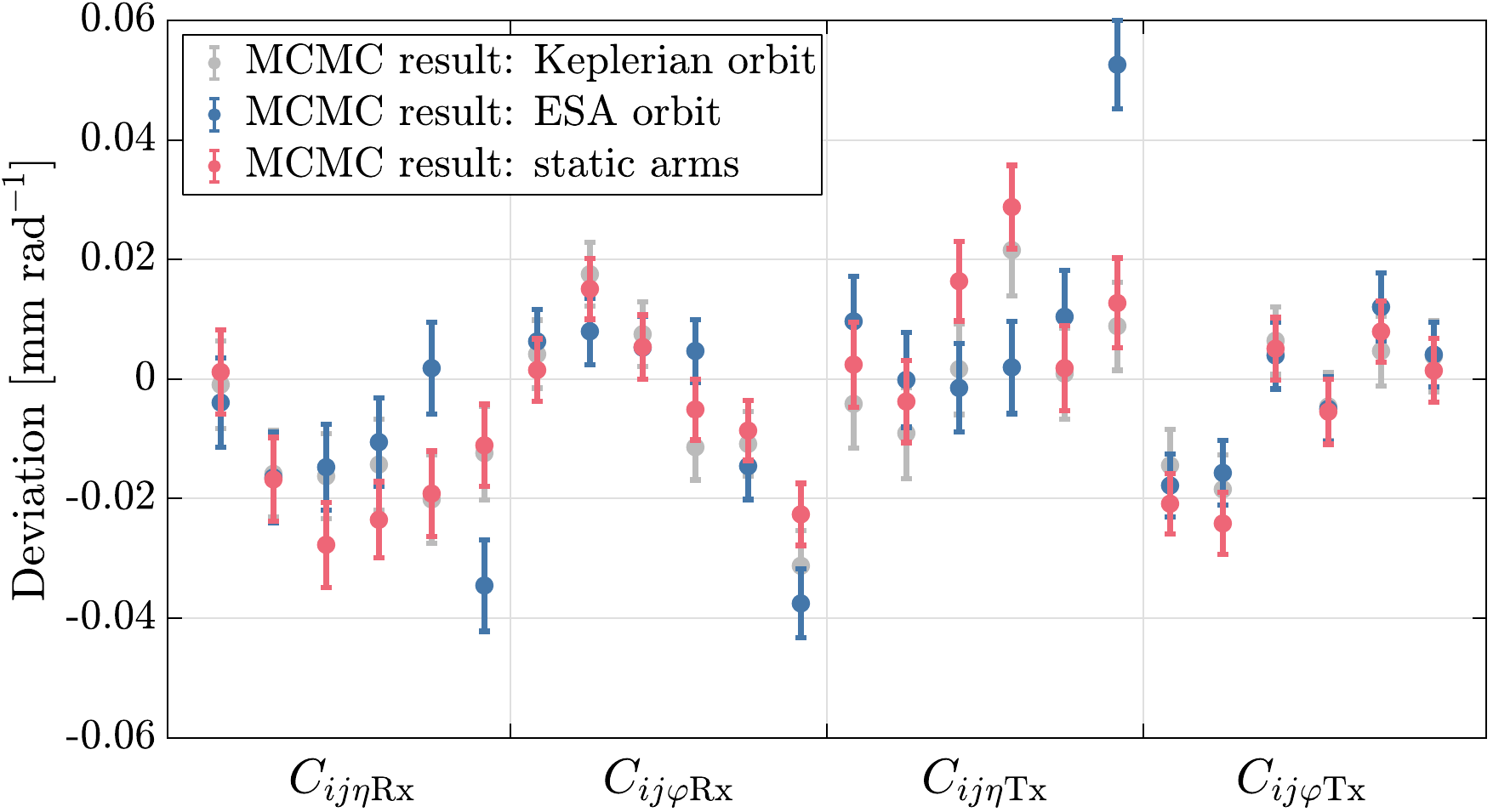} 
  \caption{Deviations of the estimated coupling coefficients from the true values for arm length changes for a LISA constellation following Keplerian (gray) or the ESA LISA science orbits (blue) and unequal static arm lengths (light-red).
  The order of the coefficient indices is $ij\in\{12,13,23,21,31,32\}$.
  The estimates considering ESA science orbits differ more from the static case than those considering Keplarian orbits, since we chose a day with almost maximal arm length changes.
  }
  \label{fig:coeff_orbit}
\end{figure}

\begin{figure}\includegraphics[width=\columnwidth]{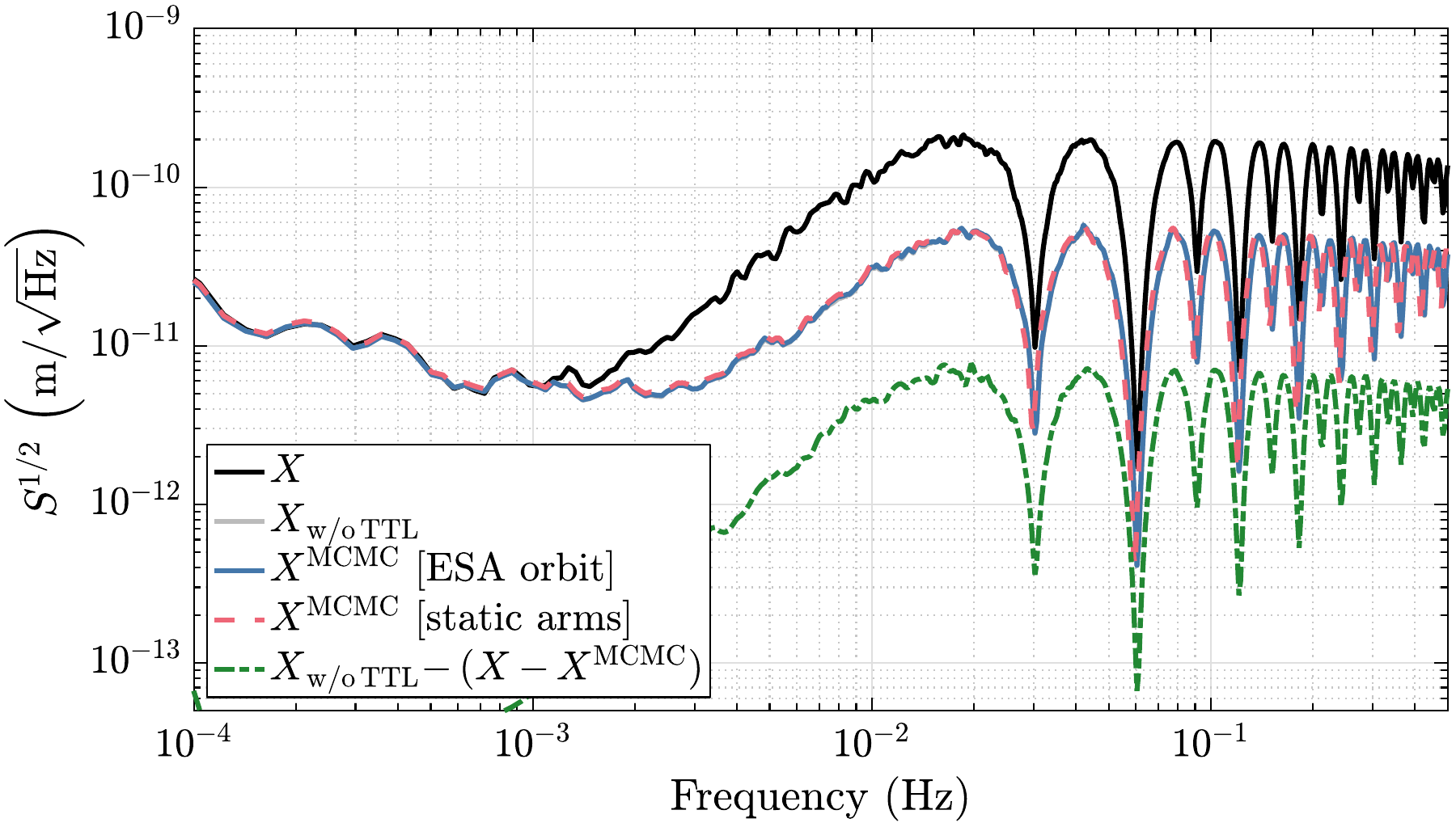} 
  \caption{ASDs of the TDI\,X combination for the simulations with (black) and without (gray) TTL coupling in the scenario with ESA science orbits.
  The blue curve shows the noise residuals after the subtraction of the TTL coupling using the MCMC fit results for the ESA science orbit case.
  For comparison, the dashed light-red curve shows the noise residuals in the case of unequal static arm lengths.
  The dash-dotted green line shows the residual TTL coupling for ESA science orbits.
  The gray curve is hardly visible since it is below the blue and light-red traces.
  }
  \label{fig:ASD_orbit}
\end{figure}

In figure~\ref{fig:coeff_orbit} we show the deviations of the estimated coupling coefficients from the true value.
We find small differences between the estimates in the different cases. All estimation errors are below 0.06\,mm/rad and therefore below the requirement of 0.1\,mm/rad.
Note that the error bars (1~$\sigma$ standard deviations) do not cover the difference between the single estimates and the true values in most cases. 
We attribute these deviations to the correlations of the coupling coefficients, which are not included in the standard deviations. The variances and covariances for the estimation method used are presented in \cite{Paczkowski2022}, along with further discussion in \cite{Paczkowski2022,Paczkowski2024}.

To demonstrate the performance of the TTL noise subtraction, we plot the residuals in figure~\ref{fig:ASD_orbit}.
We find that the residuals after subtraction in the cases of the static (light-red) and the ESA science orbit (blue) are almost identical. 
We did not plot the residual for the Keplerian orbit case since it looks like the other two residual curves.
This finding supports the assumption in \cite{Paczkowski2022} that it is sufficient to consider static unequal arm lengths in TTL coupling subtraction analyses.
We also plot the TTL noise residual (green) for the scenario with ESA science orbits. In the present paper, the term TTL noise residual describes the non-suppressed TTL noise and the DWS noise added due to the subtraction~\cite[cf.\ eq.~(18)]{Paczkowski2022}.
It is well below the other instrument noises.

In the following performance analysis with GW signals, we will use either ESA science orbits (sections~\ref{sec:GW_VB} and \ref{sec:GW_SGWB}) or assume a static constellation with equal arm lengths (sections~\ref{sec:GW_GB} and \ref{sec:GW_MBHB}).

%% file: content/TTL_GW.tex
\section{TTL coupling estimation considering gravitational wave signals}
\label{sec:GW}

In this section, we show that we can successfully suppress the TTL noise in data containing GW signals. 
This is an important case that was missing in the previous studies~\cite{Paczkowski2022,Paczkowski2024}. 
In fact, we expect to have the GW response from 
millions of sources in the LISA band at the same time~\cite{Redbook2024}.
Some sources will have a signal-to-noise ratio too small to be visible in a day's data (e.g.\ stochastic GW response). Some sources will accumulate spectral signal strength over time (e.g.\ galactic binaries). Other events will produce a short large GW response (e.g.\ massive black hole mergers).
In this paper we will focus on five different signal cases. 
First, we verify the subtraction efficiency in the presence of two verification binary signals. 
Second, we include the stochastic GW background (SGWB) from stellar origin binary black holes. 
Third, we study the TTL subtraction performance given signals of millions of galactic white dwarf binaries.
Fourth, we consider two data segments including a massive black hole binary merger.
Finally, we repeat our analysis for a data set with multiple GW sources.
For the first two cases, we used the LISA GW Response python package \cite{LISAgwresponse} to generate the LISA data sets and considered ESA LISA science orbits \cite{ESAorbits}.
In the other cases we used the `Sangria' data set from the LISA Data Challenge (LDC) 2a \cite{LDCdata,LDCsoftware}. 
As these data sets were created for equal arm length orbits, we simulated an orbit file with these properties using LISAConstants \cite{LISAconstants} and LISAOrbits \cite{LISAorbits}.
We then passed the strain Sangria data (for single or all GW source types) and the new orbit file to LISA GW Response, which computed the data file readable by LISANode.
We assumed the same noise and jitter realizations in the LISANode simulations as before.

\subsection{Verification binaries}
\label{sec:GW_VB}

As an initial test of the TTL coupling coefficient estimation in the presence of GW signals, we include the response of two verification binaries.
The signal characteristics of such binary systems are well studied and can be easily distinguished from the TTL coupling noise due to their distinct response at a single frequency. 
Therefore, we use them to verify our correct inclusion of the signals in the simulated data.
We chose the verification binaries HMCnc and ZTFJ1539 for our analysis. 
Their gravitational wave frequencies are 6.220\,mHz (HMCnc) and 4.822\,mHz (ZTFJ1539)~\cite{Kupfer2024}. 
While most of the verification binaries listed in~\cite{Kupfer2024} enter the measurement band below the frequency range where TTL coupling is expected to dominate, the selected binaries lie in the frequency regime considered in our fit.

\begin{figure}
\includegraphics[width=\columnwidth]{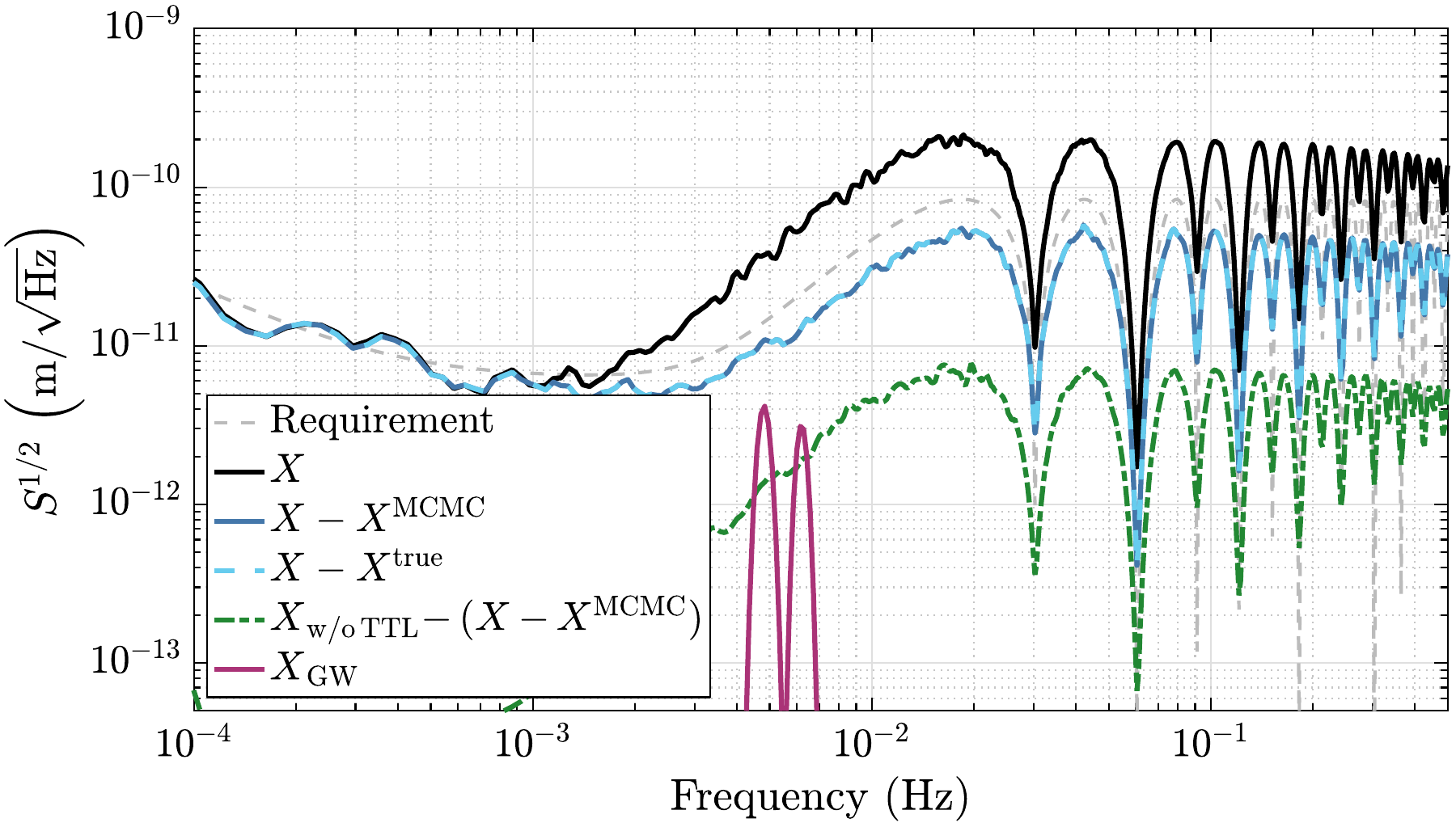} 
  \caption{Performance of the TTL coupling subtraction in the scenario with verification binaries.
  The ASD of the TDI\,X combination for the simulations with TTL coupling is plotted in black. 
  The dark-blue curve shows the ASD of the noise residual after the subtraction of the TTL coupling using the MCMC fit results and the light-blue dashed curve for the subtraction using the true coupling coefficients.
  The residual TTL coupling is plotted in dash-dotted green.
  The ASD of the GW response in TDI\,X is shown in purple.
  All ASDs were computed for one day measurements at 2\,Hz.
  The dashed gray line corresponds to the LISA requirement.
  }
  \label{fig:ASD_VB}
\end{figure}

\begin{figure}
\includegraphics[width=\columnwidth]{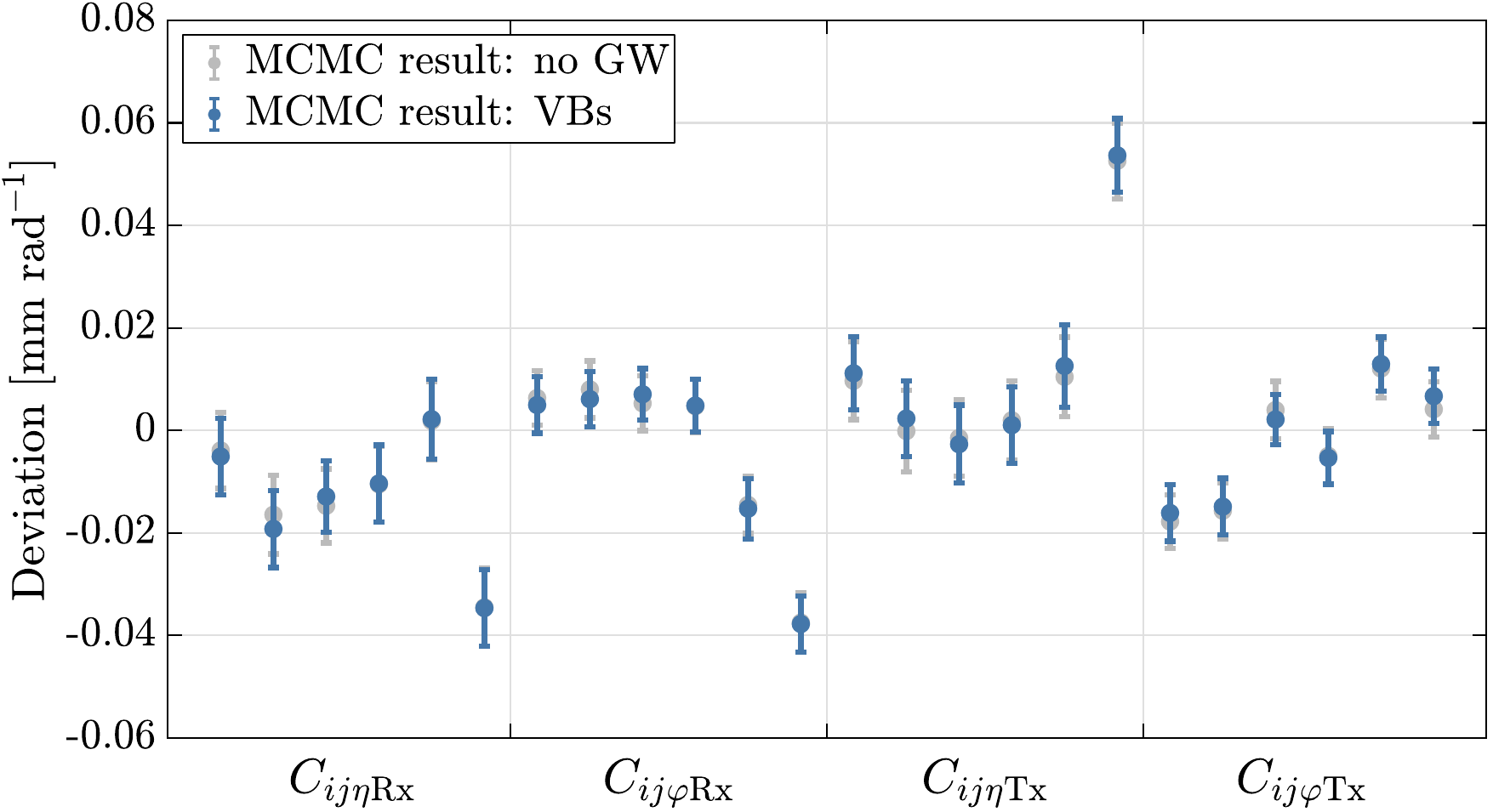} 
  \caption{Deviations of the estimated coupling coefficients from the true values with and without GW signals. 
  Blue: Simulated data with VB signals. 
  Gray: Simulated data without GW signals. 
  In both simulations, we used ESA science orbits.
  The order of the coefficient indices is $ij\in\{12,13,23,21,31,32\}$.
  }
  \label{fig:coeff_VB}
\end{figure}

For the time interval of a single day, the signal from the verification binaries is well below the noise sources, see figure~\ref{fig:ASD_VB}.
Therefore, the fitted coefficients are almost identical to the case without GW signals and deviate less than 0.06\,mm/rad from the true values, see figure~\ref{fig:coeff_VB}.
Also, the subtraction of the TTL coupling works well (figure~\ref{fig:ASD_VB}). The noise level after subtraction lies significantly below the LISA mission noise requirement, which was derived from the TM acceleration and displacement noise, see~\cite{Paczkowski2022}. 
The noise residuals using the fitted and the true coupling coefficients show no noticeable difference. In fact, their difference is one order of magnitude smaller than the plotted residual noise curves.

\subsection{Stochastic gravitational wave background from stellar-origin black hole binaries}
\label{sec:GW_SGWB}

\begin{figure}
\includegraphics[width=\columnwidth]{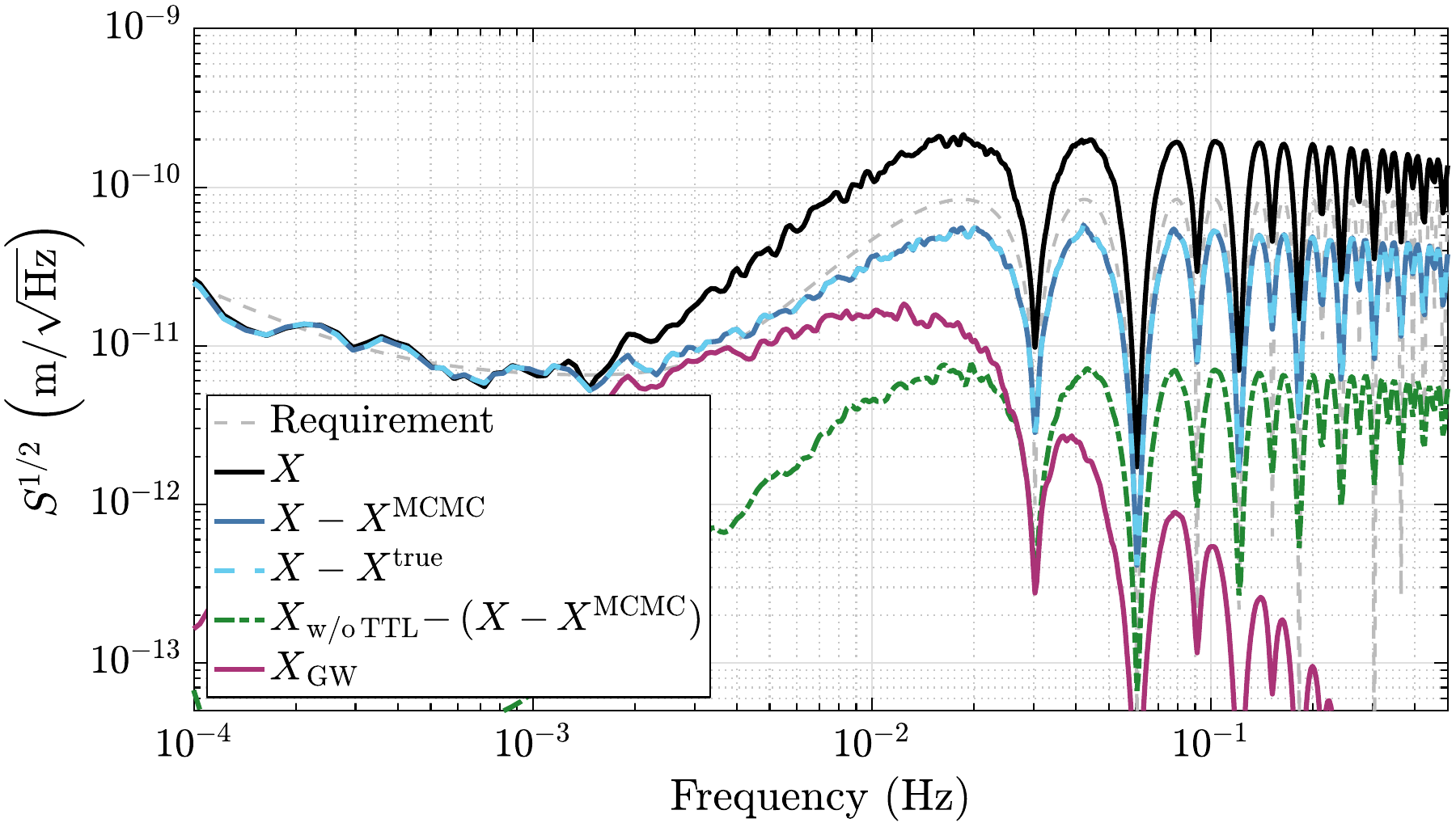} 
  \caption{Performance of the TTL coupling subtraction in the scenario with SGWB.
  We plot the same selection of curves as in figure~\ref{fig:ASD_VB}.
  }
  \label{fig:ASD_SGWB}
\end{figure}

\begin{figure}
\includegraphics[width=\columnwidth]{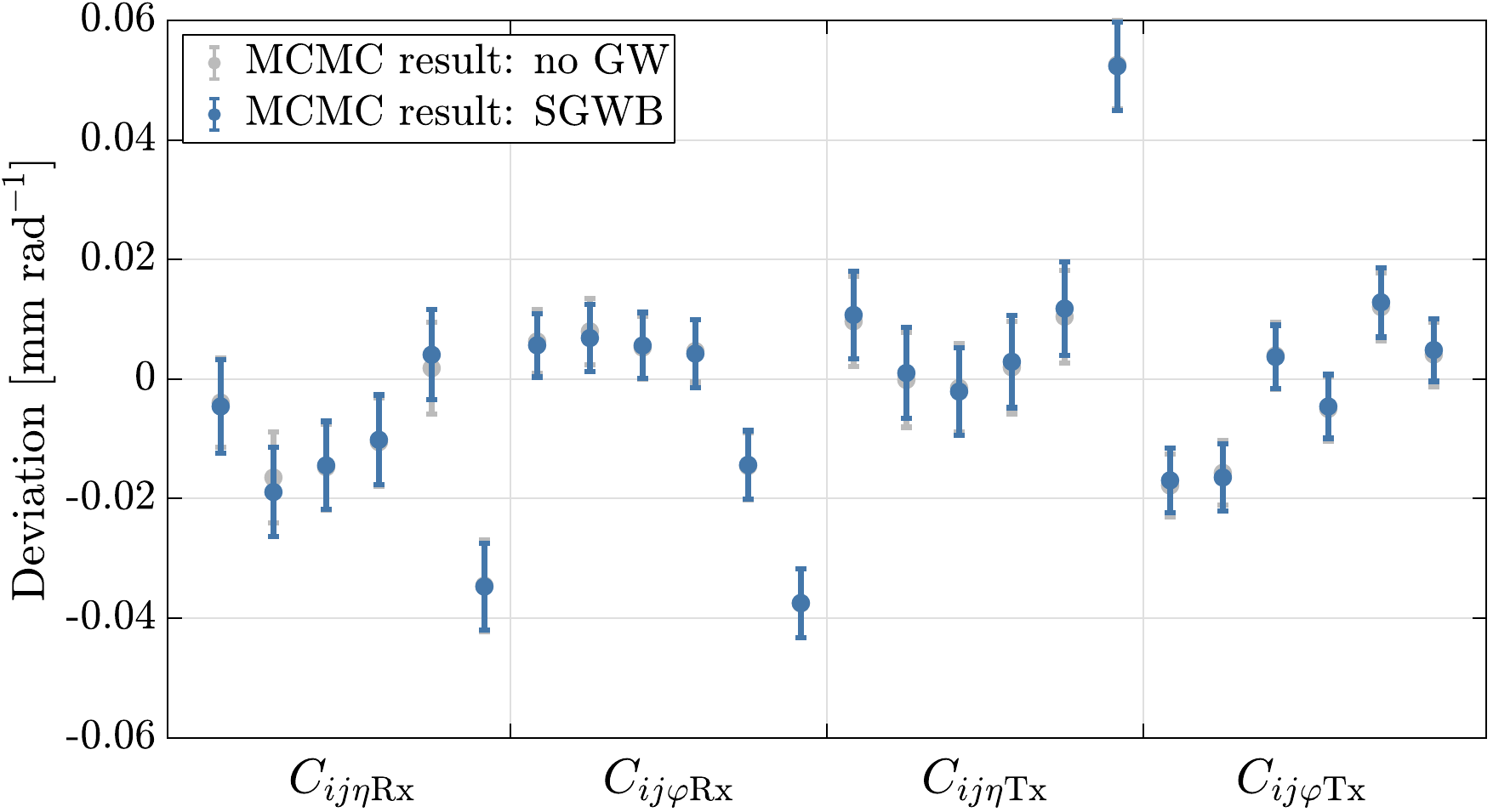} 
  \caption{Deviations of the estimated coupling coefficients from the true values for data sets with (blue) and without (gray) SGWB.
  The order of the coefficient indices is $ij\in\{12,13,23,21,31,32\}$.
  }
  \label{fig:coeff_SGWB}
\end{figure}

In our second test, we consider the response of the SGWB from stellar-origin mass black hole binaries. 
The SGWB are GW signals that are best characterized statistically \cite{LVK-3OR}.
The distinction of SGWB from instrument noise is therefore the subject of several ongoing studies, e.g.\ \cite{Muratore2024,Hartwig2023}.
There exist two types of SGWB: astrophysical and cosmological SGWB. 
Astrophysical SGWB describes the superposition of GW signals from many unresolved astrophysical sources including black hole and neutron star binaries \cite{LVK-3OR,Auclair2023}.
Its largest contributors are the white dwarf binaries, that we will discuss in the subsection below considering constant equal arms.
As a pretest, we investigate here the TTL subtraction scheme for a stellar-origin binary black hole dominated SGWB and flexing LISA arms defined by the ESA LISA science orbits.
We use the model from \cite{Babak2023}, which describes the background by the power law
\begin{eqnarray}
  h^2\Omega_{\rm GW}(f) = h^2\Omega_{\rm GW}(f_\ast) \left(\frac{f}{f_\ast} \right)^{2/3} \,,
\end{eqnarray}
where $f_\ast$ is an arbitrary pivot frequency.
We set $h^2\Omega_{\rm GW}(f_\ast=3\,\rm mHz)=1.15\cdot10^{-12}$, which is the maximum of the range computed in \cite{Babak2023}.
The resulting GW response for one day of measurement time is shown in figure~\ref{fig:ASD_SGWB} (purple) together with the instrument noise including TTL coupling (black).
The SGWB signal remains below the TTL coupling and the other instrument noise sources.
Therefore, the TTL coupling coefficients can be well fitted (figure~\ref{fig:coeff_SGWB}). In fact, the effect of the SGWB on the coefficient estimation is marginal.
The noise residual after subtraction using the fit coefficients suppresses the measurement noise below the LISA requirement (blue curve in figure~\ref{fig:ASD_SGWB}) and the TTL noise residual is as small as in the previously investigated cases (compare green curves in figures~\ref{fig:ASD_orbit}, \ref{fig:ASD_VB} and \ref{fig:ASD_SGWB}).

\subsection{Detached white dwarf galactic binaries}
\label{sec:GW_GB}

\begin{table}
\setlength{\tabcolsep}{1.5ex}
\caption{Properties of the two 1-day data segments with detached galactic binaries from the Sangria data set: Start time $t_0$ and the RMS of the GW strains during that day.}
\begin{tabular}{@{}lcc}
\toprule
  & $t_0$ & strain RMS \\
\midrule
Data segment 1 (GB1)
  & 10\,s        & 4.8$\cdot10^{-22}$ \\
Data segment 2 (GB2) 
  & 10303235\,s  & 5.9$\cdot10^{-22}$ \\
\bottomrule
\end{tabular}
\label{tab:sangria_gb}
\end{table}

Here, we study the GW response %
from the Sangria data set \cite{LDCdata}. The full data set contains about 30\,million white dwarf binaries splitting into detached binary systems that evolve under gravitational radiation reaction and interacting binaries with a steady mass transfer. Furthermore, the data set includes verification binaries and black hole mergers \cite{LDCdoc}. 
In this section, we use only the GW data from the detached binary system, which build the majority of the white dwarf binaries in the Sangria data set.
As we show below, these binary signals are louder than the GW signals in the two previous studies. They exceed the instrument noise at frequencies considered for the TTL coefficient estimation.
The galactic binary signal response depends on the orientation of the LISA constellation and thus changes with time. Therefore, we investigate here two data segments for two different times of the mission. Their properties are summarized in table~\ref{tab:sangria_gb}.
There, we computed the root mean square (RMS) of the GW strain for both periods in order to provide the signal strength with a number.
It can be seen that the signal response for the second data segment is stronger than for the first one. 
In the following, we show the results for both cases separately for the sake of clarity.

\begin{figure}
\includegraphics[width=\columnwidth]{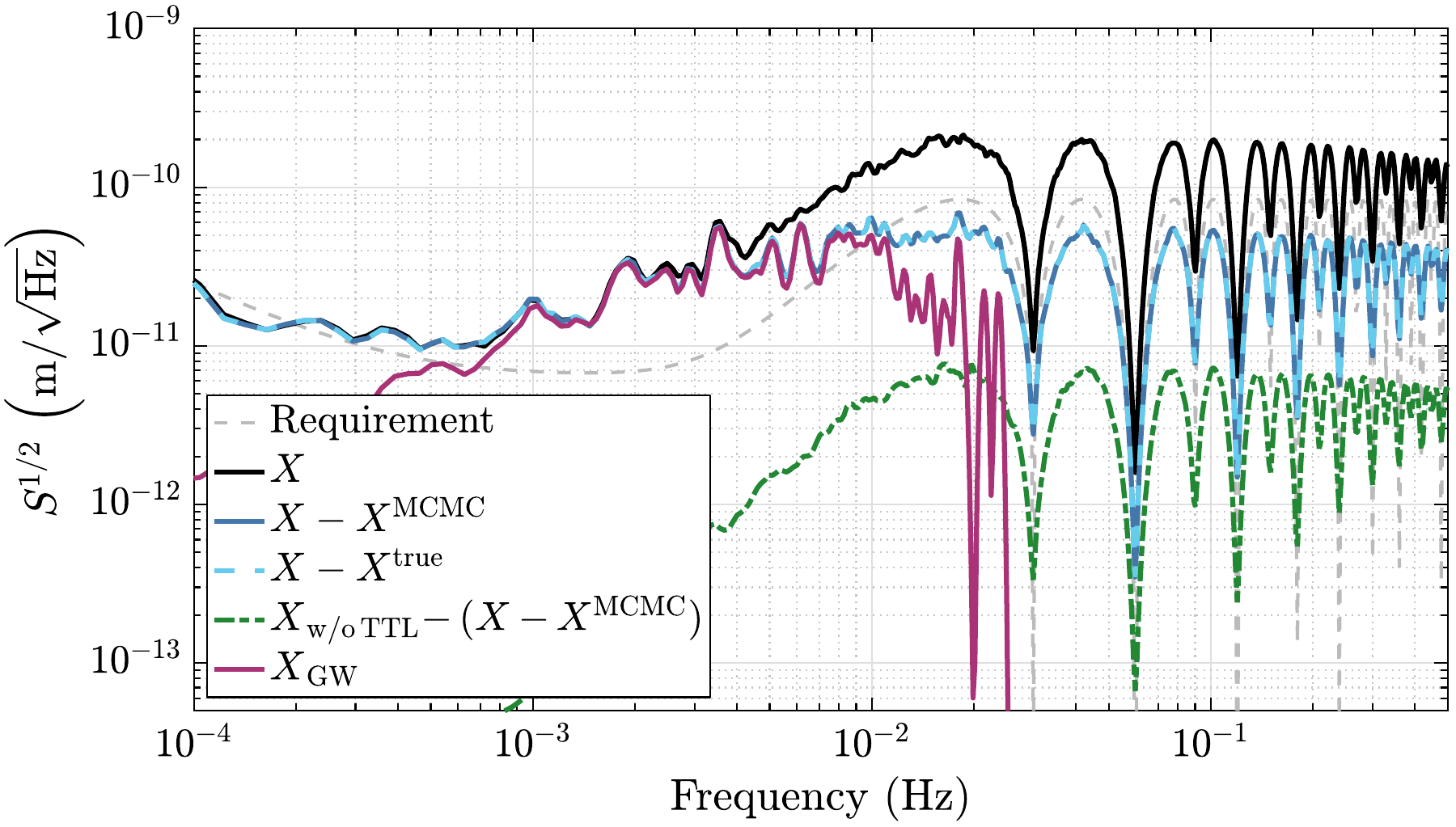} 
  \caption{Performance of the TTL coupling subtraction in the case GB1. %
  We plot the same selection of curves as in figure~\ref{fig:ASD_VB}.
  In this scenario, the GW signal (purple) dominates the measurement before (black) and after the TTL noise subtraction (two shades of blue). The TTL residual (dash-dotted green) remains unchanged compared to figure~\ref{fig:ASD_VB}.
  }
  \label{fig:ASD_sangria_GB}
\end{figure}

\begin{figure}
\includegraphics[width=\columnwidth]{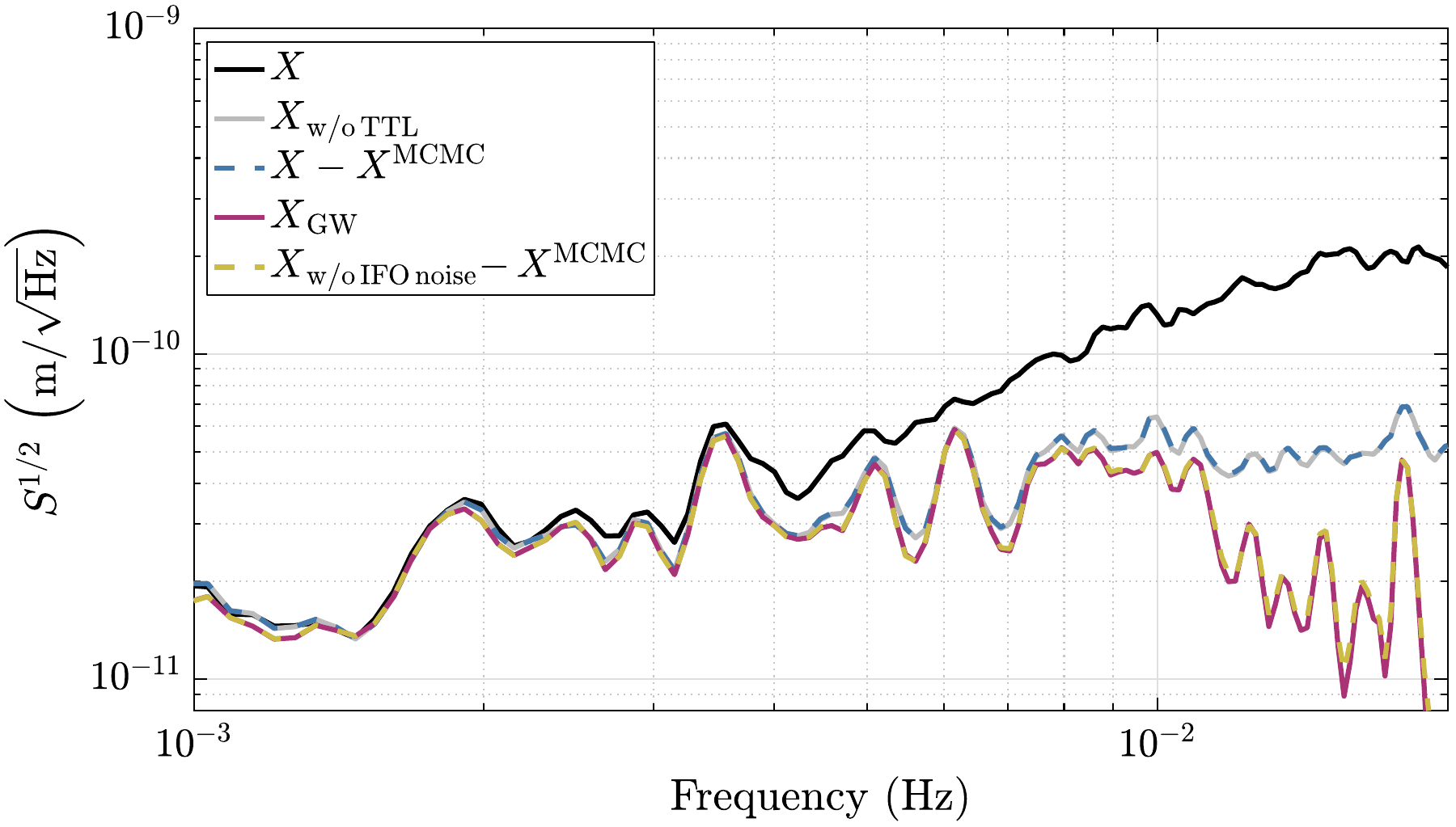} 
  \caption{Close-up view on the performance of the TTL coupling subtraction in the case GB1. %
  The ASD of the TDI\,X combination for the simulations with (without) TTL coupling is plotted in black (gray). 
  The dashed dark-blue curve shows the ASD of the residual after the subtraction of the TTL coupling using the MCMC fit results.
  The ASD of the GW response after TDI is shown in purple.
  The ASD of the residual after subtracting the TTL model with the fitted MCMC coefficients from data without instrument noise is plotted in dashed yellow.
  All ASDs were computed for one day measurements at 2\,Hz.
  }
  \label{fig:ASD_sangria_GB_zoom}
\end{figure}

First we look at the performance of the TTL subtraction for data segment GB1.
The ASDs of the noise residuals are plotted in figure~\ref{fig:ASD_sangria_GB}.
The GW signal response exceeds the noises including TTL coupling (blacks) at frequencies below 10\,mHz.
The subtraction of the fitted TTL coupling model (dark-blue) suppresses the residual to the level of the GW signal, where it is dominant, and below the LISA requirement otherwise. 
We see no notable difference in the performance using the fitted coefficient estimates or the true coupling coefficients (compare the dark- and light-blue traces in figure~\ref{fig:ASD_sangria_GB}).
Also, the TTL coupling residual after subtraction (dash-dotted green) is about one order of magnitude below the other noise sources.
This TTL coupling residual was added to the other instrument noises when subtracting the TTL noise model. In detail, we add in noise due to the coefficient estimation error but also the DWS readout noise, which is scaled by the coupling coefficients, compare equations~\eqref{eq:DWS} and \eqref{eq:TTL_singlelink} or \cite[cf.\ eq.~(18)]{Paczkowski2022}.
Therefore, we review the residuals at low frequencies, where the GW signal is dominant, see figure~\ref{fig:ASD_sangria_GB_zoom}.
For our settings, we find that the residual after TTL noise suppression (dark-blue) overlaps with the simulated data without TTL coupling (gray) for the given level of DWS readout noise. In particular, this shows that the residual after TTL noise subtraction is dominated by the GW signal and by other instrument noises. The residual TTL noise including the added DWS readout noise is much smaller as shown in figure~\ref{fig:ASD_sangria_GB}.
Furthermore, we compare the pure GW signal response (purple) with the residual found when subtracting the fitted TTL noise model from simulated data with no instrument noise other than TTL coupling (yellow). 
Both curves align well with each other. 
Therefore, we do not observe a degradation of the GW signal due to the TTL subtraction process.

\begin{figure}
\includegraphics[width=\columnwidth]{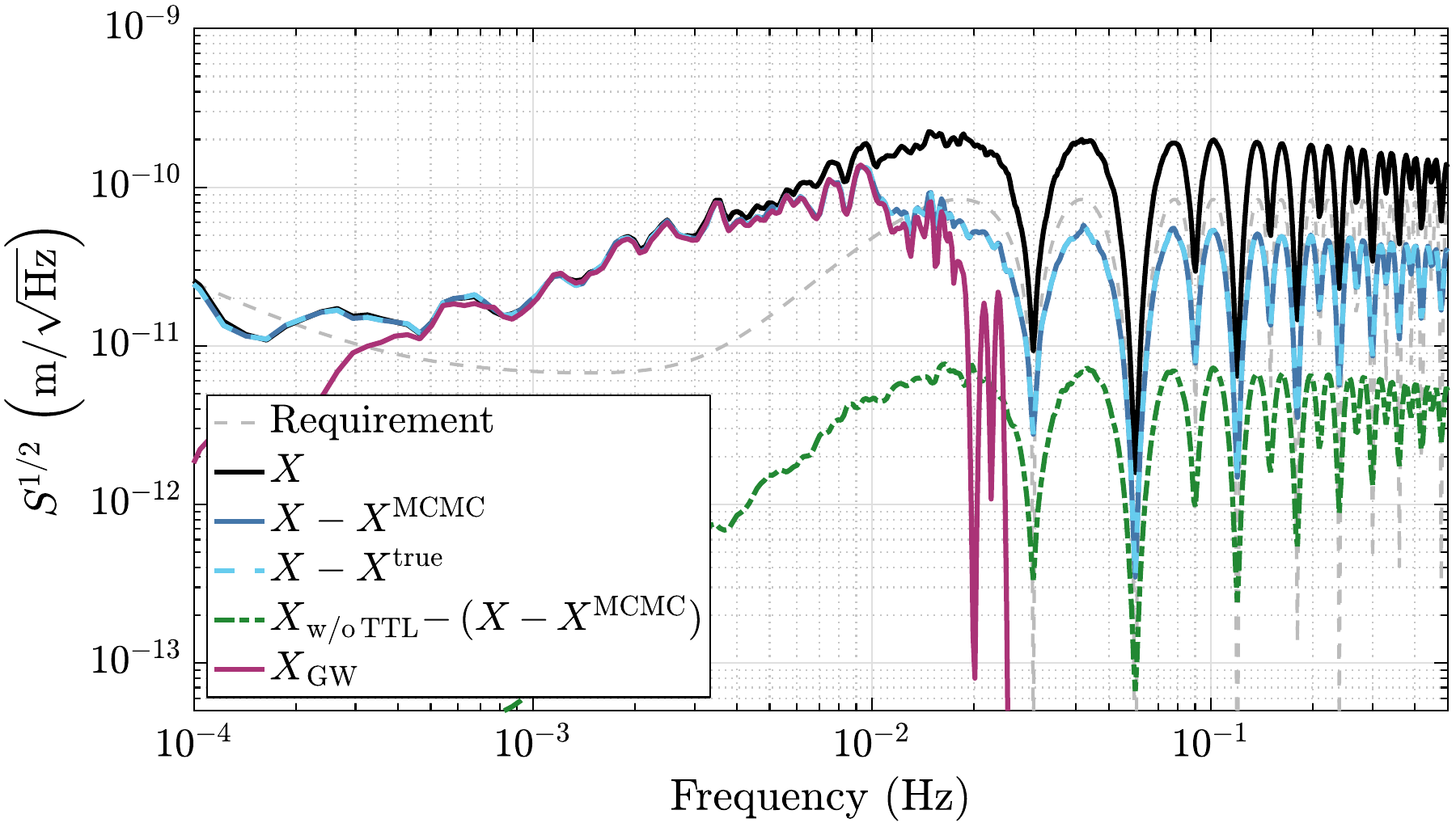} 
  \caption{Performance of the TTL coupling subtraction in the case GB2. %
  We plot the same selection of curves as in figure~\ref{fig:ASD_VB} (and figure~\ref{fig:ASD_sangria_GB}).
  }
  \label{fig:ASD_sangria_GBmax}
\end{figure}

We repeat our analysis for the detached galactic binary response in data segment GB2. %
The signal and the performance of the TTL coupling subtraction are shown in figure~\ref{fig:ASD_sangria_GBmax}. 
We find that the signal response is stronger and of similar shape compared to the previously analyzed galactic binary case. 
The TTL coupling subtraction suppresses the noise below the LISA requirement at frequencies, where noise dominates the GW signals, and the TTL noise residual (green) is as small as before.

\begin{figure}
\includegraphics[width=\columnwidth]{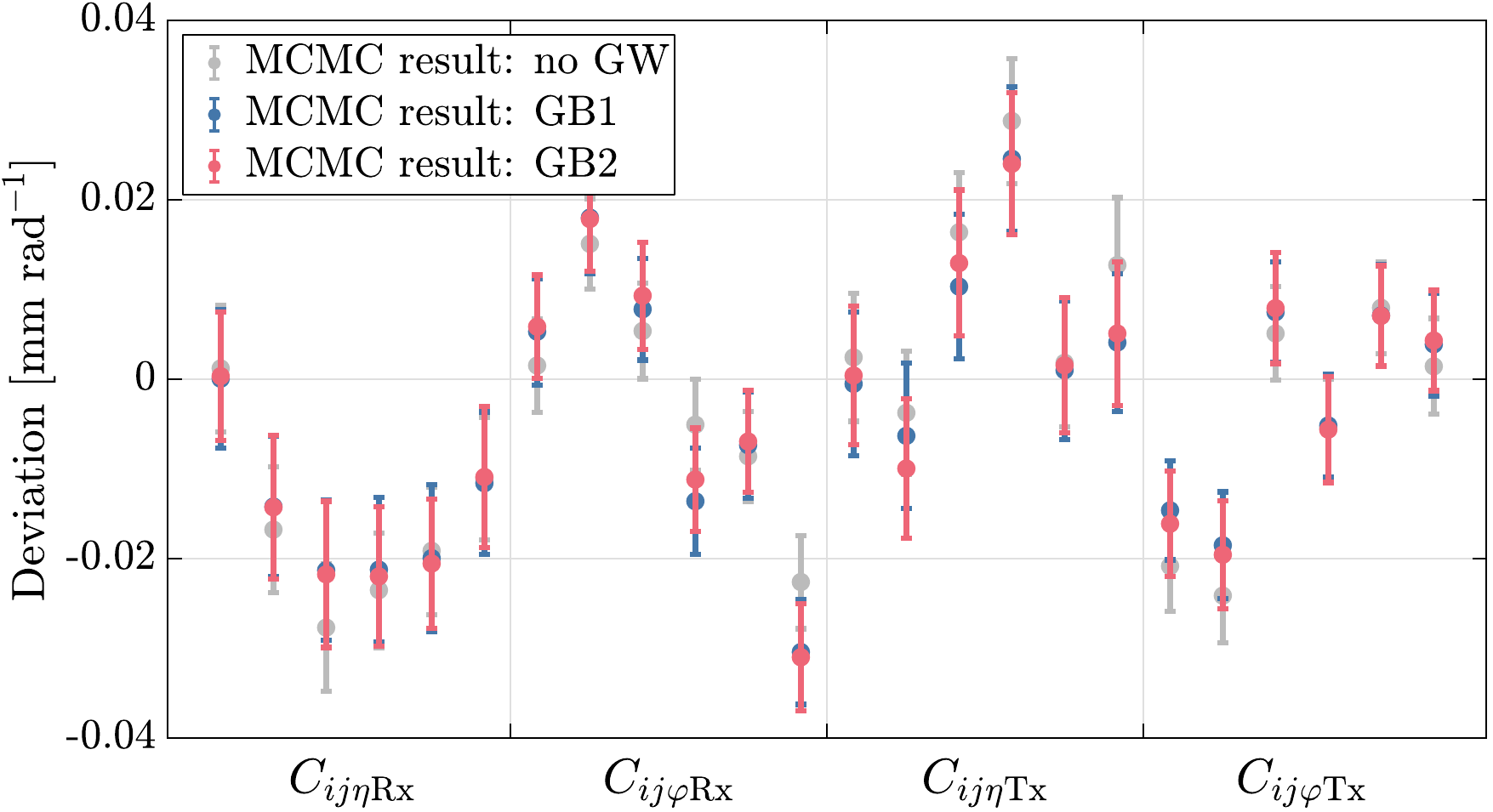} 
  \caption{Deviations of the estimated coupling coefficients from the true values with and without GW signals. 
  Blue: Simulated data with GW signals (GB1). %
  Light-red: Simulated data with GW signals (GB2). %
  Gray: Simulated data without GW signals considering static equal arm length orbits.
  The order of the coefficient indices is $ij\in\{12,13,23,21,31,32\}$.
  }
  \label{fig:coeff_sangria_gb}
\end{figure}

Lastly, we revisit the estimated TTL coupling coefficients for both galactic binary scenarios.
These are plotted in figure~\ref{fig:coeff_sangria_gb}.
We find that the estimation errors are below the 0.1\,mm/rad requirement and agree within the error bars with the estimates found in the case where no GW signals were considered.

\subsection{Massive black hole binary mergers}
\label{sec:GW_MBHB}

\begin{table*}
\setlength{\tabcolsep}{2ex}
\caption{Properties of the two data segments with MBHB mergers from the Sangria data set: Start time $t_0$, time of coalescence $t_c$, masses $M_1$ and $M_2$ of the two merging black holes in units of solar masses $M_\odot$, distance $d$ and the redshift $z$.}
\begin{tabular}{lcccccc}
\toprule
  & $t_0$ & $t_c$ 
  & $M_1$ & $M_2$ 
  & $d$ & $z$ \\
\midrule
Data segment 1 (MBHB1)
  & 13.57$\times10^6$\,s  & 13.62$\times10^6$\,s
  & 3.20$\times10^6$\,$M_\odot$ & 3.06$\times10^6$\,$M_\odot$
  & 24.83\,kly  & 2.88 \\
Data segment 2 (MBHB2) 
  & 11.22$\times10^6$\,s  & 11.26$\times10^6$\,s
  & 9.16$\times10^5$\,$M_\odot$ & 7.02$\times10^5$\,$M_\odot$
  & 7.71\,kly   & 1.10 \\
\bottomrule
\end{tabular}
\label{tab:sangria_MBHB}
\end{table*}

Next, we investigated TTL noise subtraction in the case of massive black hole binary (MBHB) mergers.
The Sangria dataset contained 15 such mergers.
Again, we picked two time segments with one merger signal each set to be roughly in the middle of a one-day simulation.
The data segment and merger properties are provided in table~\ref{tab:sangria_MBHB}.

\begin{figure}
\includegraphics[width=\columnwidth]{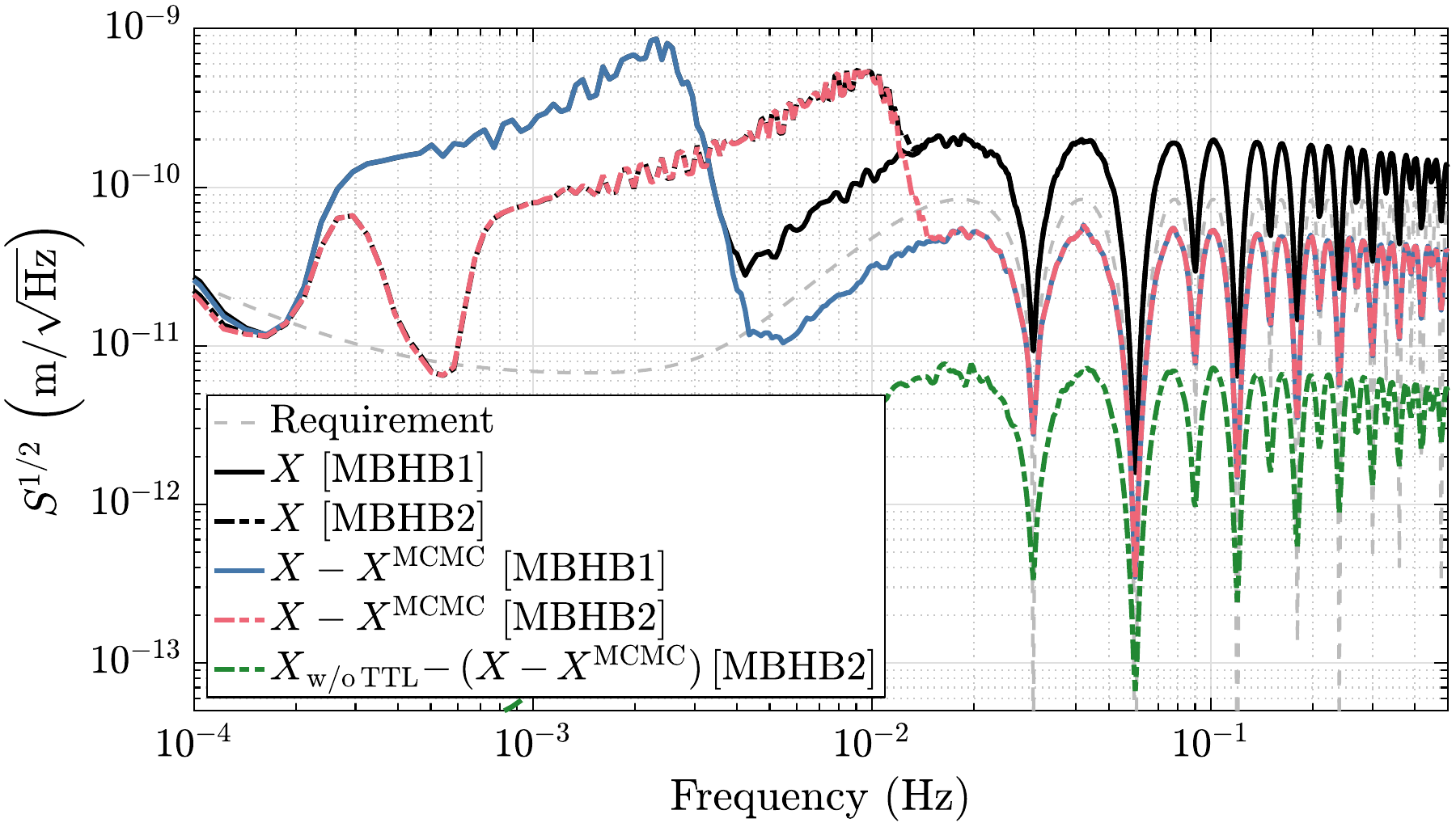} 
  \caption{Performance of the TTL coupling subtraction for data containing MBHB mergers.
  The ASDs of the TDI\,X combination for the simulations with TTL coupling is plotted in black. 
  The dark-blue and light-red curves show the ASD of the residual after the subtraction of the TTL coupling using the MCMC fit results. %
  The dash-dotted green curve is the TTL noise residual for the case MBHB2.
  See table~\ref{tab:sangria_MBHB} for the reference of the two GW cases.
  The dashed gray line corresponds to the LISA requirement.
  }
  \label{fig:ASD_sangria_MBHB}
\end{figure}

\begin{figure}
\includegraphics[width=\columnwidth]{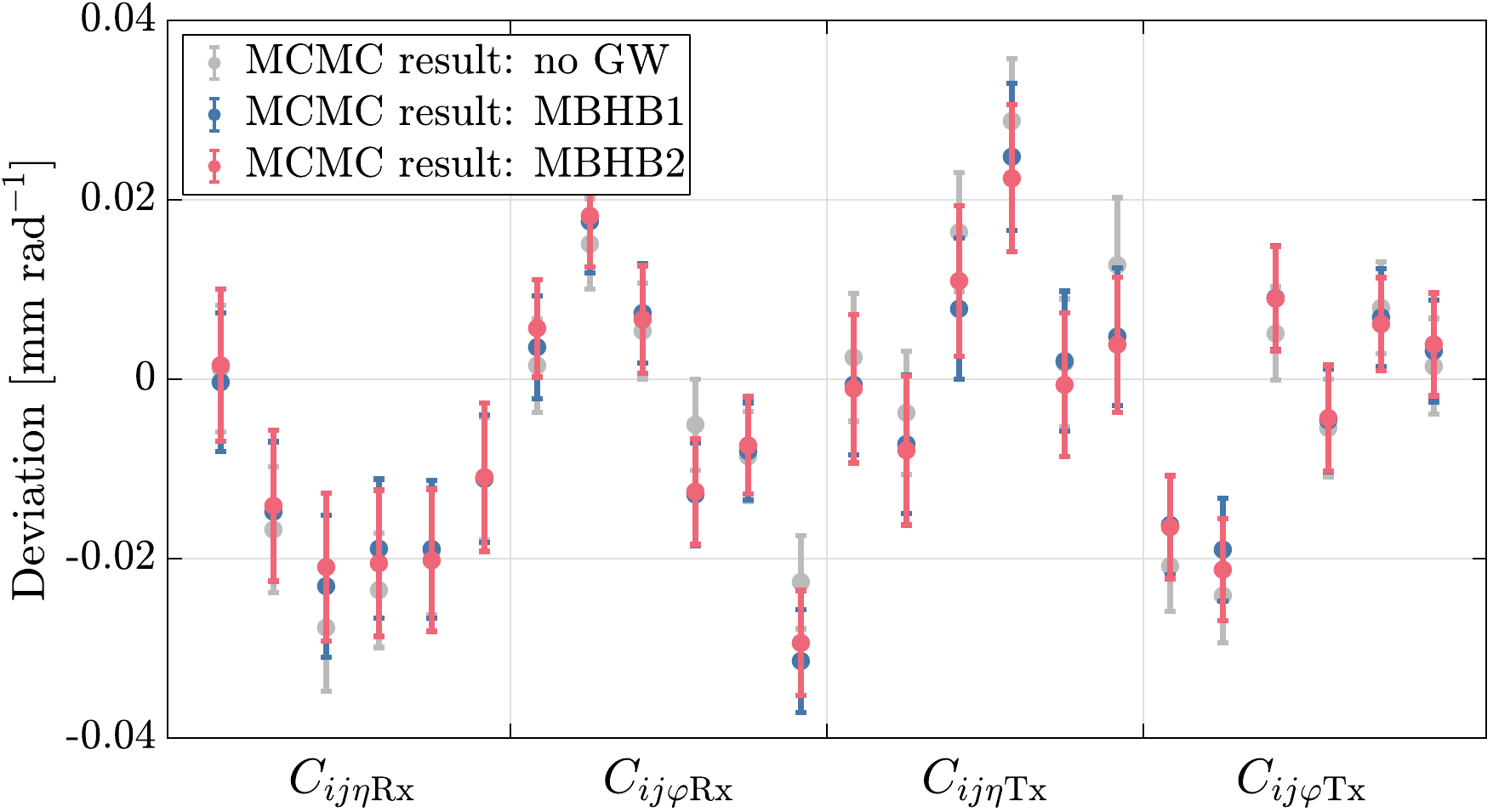} 
  \caption{Deviations of the estimated coupling coefficients from the true values with and without GW signals. 
  Blue: Simulated data for the case MBHB1. 
  Light-red: Simulated data for the case MBHB2.
  See table~\ref{tab:sangria_MBHB} for the reference of the two GW cases.
  Gray: Simulated data without GW signals considering static equal arm length orbits.
  The order of the coefficient indices is $ij\in\{12,13,23,21,31,32\}$.
  }
  \label{fig:coeff_sangria_MBHB}
\end{figure}

\begin{figure}
\includegraphics[width=\columnwidth]{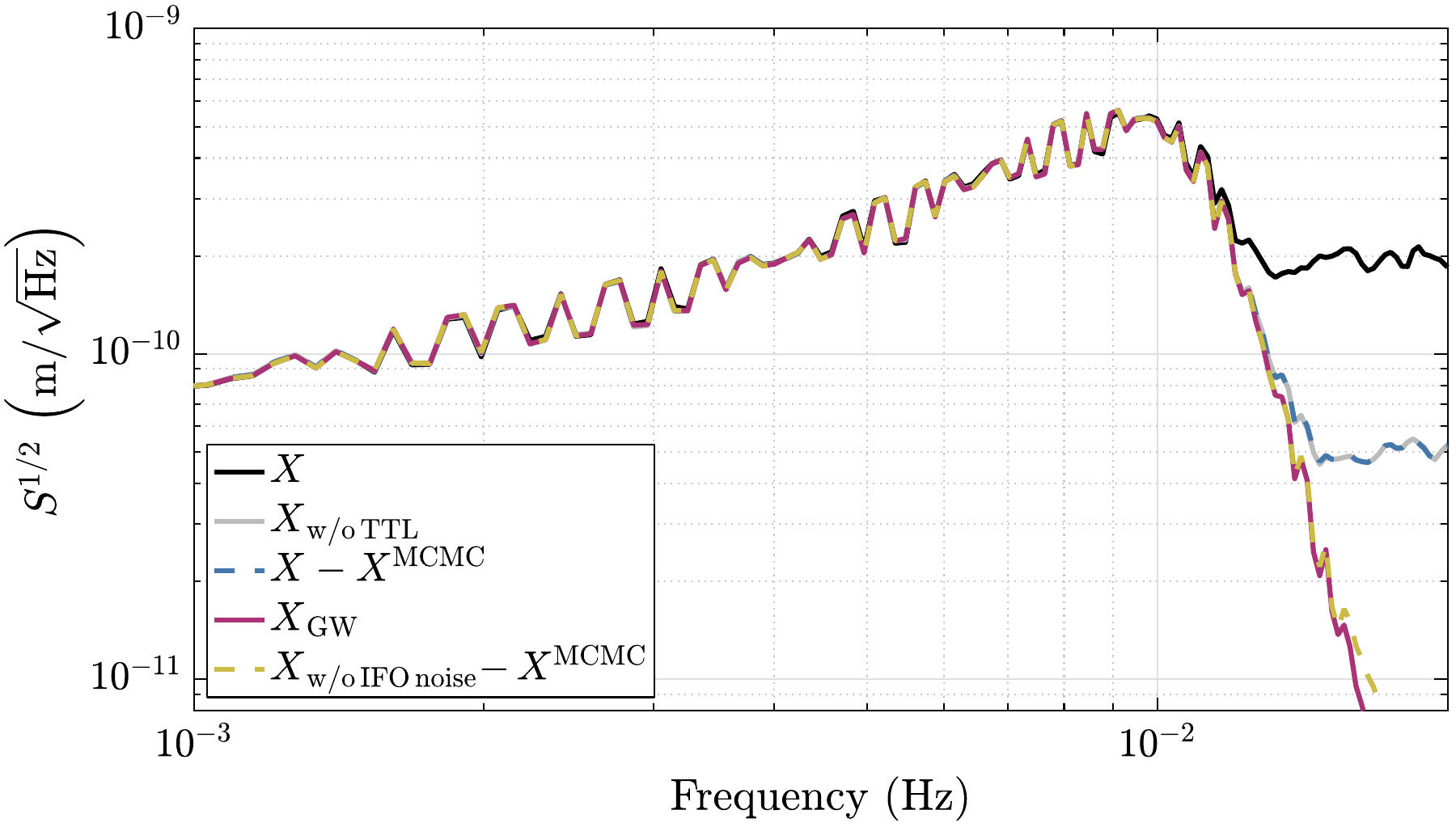} 
  \caption{Close-up view on the performance of the TTL coupling subtraction for the case MBHB2.
  We plot the same selection of curves as in figure~\ref{fig:ASD_sangria_GB_zoom}.
  }
  \label{fig:ASD_sangria_MBHB_zoom}
\end{figure}

In figure~\ref{fig:ASD_sangria_MBHB} we see the signal responses for the two days with one merger each. The response maxima are above 2\,mHz and around 10\,mHz. The signals exceed the TTL coupling below 4\,mHz or 12\,mHz respectively. %
The subtraction of the TTL noise using the MCMC fit results suppresses this noise well below the requirement at frequencies not dominated by the merger signals.
When fitting the TTL coupling noise, we find that the errors of the estimated coefficients are below 0.04\,mm/rad for both merger cases and therefore well below the requirement of 0.1\,mm/rad, as can be seen in figure~\ref{fig:coeff_sangria_MBHB}.
Exemplarily, we also show the close-up view on the performance of the TTL coupling subtraction for the case MBHB2 in figure~\ref{fig:ASD_sangria_MBHB_zoom}.
The signal of this merger has a considerable overlap with the frequency range where TTL coupling is the dominant noise source (after TDI). Therefore, the TTL noise needs to be well subtracted.
The residual after the subtraction of the TTL noise using the fitted MCMC coefficients (blue) approximates well the noise level in the data set without TTL coupling (gray). 
To confirm that we do not alter the GW signal by subtracting TTL noise, we also plot the residual using the same fitted TTL noise model, but subtracting it from a data set with no instrument noises except for TTL coupling (yellow).
This residual matches the GW signal curve (purple), confirming that the TTL noise subtraction does not perceivably alter the GW response in our simulation.

It appears that the MBHB signals had only a minor effect on the coefficient estimate or the quality of the TTL noise subtraction from the data (compare also green curve in figure~\ref{fig:ASD_sangria_MBHB}), although they were to be found in the frequency range of the fit. Note that there is still a sizable part of the frequency band where the TTL coupling is dominant.
Therefore, we repeated the analysis for the fit frequency range from 3\,mHz to 200\,mHz. The results are shown in appendix~\ref{app:GW_200mHz}. We find a comparable accuracy of the estimated coupling coefficients.
Still, the MBHB results should be treated with caution. 
In our analysis, the jitter shapes are white in the frequency range of the fit.
While the jitter levels refer to the LISA performance model, the white noise shape is not expected.
During the LISA mission, we expect the jitter shapes to be colored. We cannot say, whether or how other jitter shapes would affect the findings of this paper.
An investigation of the TTL coefficient estimation accuracy for colored jitter and different MBHB mergers has been presented in~\cite{Hartig2024}.
The here presented results cannot directly be compared with~\cite{Hartig2024} since different MBHB merger characteristics have been used in~\cite{Hartig2024}. However, it appears that the time domain least squares estimator used in that work is more sensitive to merger signals than our MCMC method.

\subsection{Multiple GW sources}
\label{sec:GW_Sangria}

\begin{figure}
\includegraphics[width=\columnwidth]{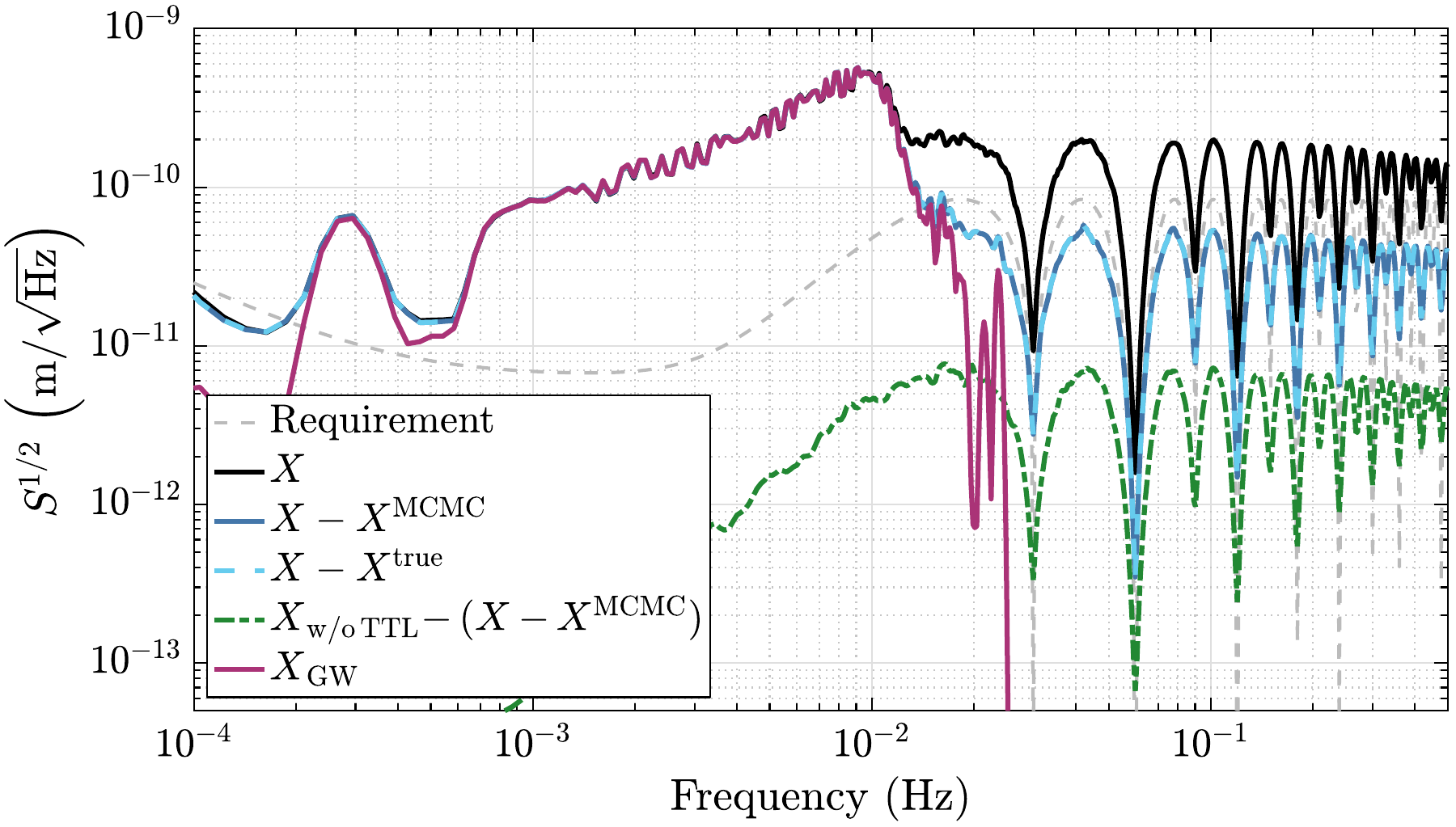} 
  \caption{Performance of the TTL coupling subtraction in the case of the full Sangria data set.
  We plot the same selection of curves as in figure~\ref{fig:ASD_VB}.
  }
  \label{fig:ASD_sangria}
\end{figure}

\begin{figure}
\includegraphics[width=\columnwidth]{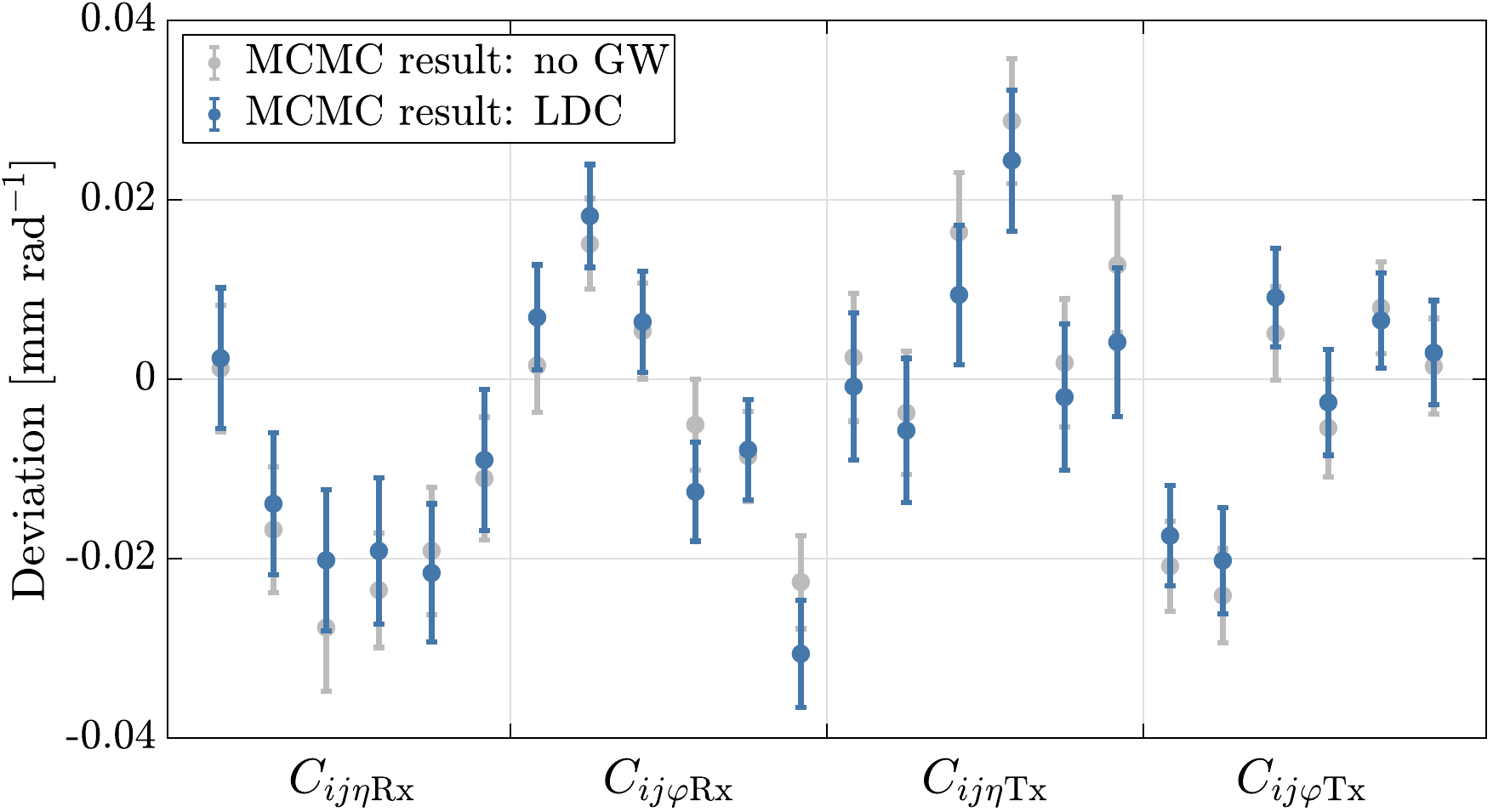} 
  \caption{Deviations of the estimated coupling coefficients from the true values for data sets with (blue) and without (gray) the GW sources of the Sangria data set from the LDC.
  The order of the coefficient indices is $ij\in\{12,13,23,21,31,32\}$.
  }
  \label{fig:coeff_sangria}
\end{figure}

Finally, we repeated our analysis considering all GW signals in the Sangria data set, namely verification binaries, interacting and detached galactic binaries, as well as black hole mergers \cite{LDCdata,LDCsoftware}.
We simulated the LISA data for the time period with the MBHB2 merger discussed above.
As expected this merger is the dominant GW source on this day, compare figures~\ref{fig:ASD_sangria_MBHB} and \ref{fig:ASD_sangria}.
Consequently, the TTL coupling coefficient estimation worked equally well (figure~\ref{fig:coeff_sangria}) and the TTL noise was efficiently minimized (figure~\ref{fig:ASD_sangria}).
Therefore, this final test case demonstrates that given our settings we can successfully fit and subtract the TTL noise in the presence of multiple different GW signals.

%% file: content/summary.tex
\section{Summary}
\label{sec:summary}

In the present paper, we have investigated the performance of the TTL coupling suppression via subtraction in post-processing for different GW signals in the simulated data.
The data were generated using the simulator LISANode, and we chose the same noise realizations in all scenarios for a better comparison of the different cases. 
We then used an MCMC algorithm to fit the TTL coupling coefficients to the TDI\,AET combinations of the interferometric length measurements.
Our analysis showed that the coupling coefficients could be fitted to within a deviation of $\pm$0.1\,mm/rad in all the scenarios.
By subtracting the TTL noise model from the data using the fitted coupling coefficients, we could suppress the total noise below the LISA mission noise requirement. The residual TTL noise was shown to be well below the other instrument noises. Due to the small deviations of the coupling coefficients, the differences of the residuals after subtraction using the fitted or the true coupling coefficients are marginal.
We also investigated the GW signals after the TTL coupling subtraction. 
We could show that neither the coefficient inaccuracies nor the DWS readout noise added significantly to the data when subtracting the TTL noise.
These results are a first evidence that the TTL suppression scheme planned for the LISA mission will not only suppress the noise well below the requirement, but will also not affect the GW signal content in the data. However, the latter was shown only indirectly in figures~\ref{fig:ASD_sangria_GB_zoom} and \ref{fig:ASD_sangria_MBHB_zoom} for one day of data.
The results would need to be confirmed considering the longer integration times for GW analyses and their ultimate effect on the astrophysical parameter estimation.
Despite the overall reassuring results of our analysis, some aspects will require further investigation in the future. 
Our simulations of the LISA data considered only the SC and MOSA jitter, which is white in the frequency range where TTL coupling is assumed to be dominant.
In reality, we expect the jitter to be colored.
For instance, the SC jitter will roll off at higher frequencies reducing the frequency range in which TTL coupling is dominant. 
We have tested the TTL coefficient fit for smaller frequency ranges than the one used in most of this paper and found comparable results (see e.g.\ appendix~\ref{app:GW_200mHz}).
It remains to be shown that the GW signals and the jitter can also be well distinguished for other jitter shapes.
The implementation of the LISA dynamics and control loops \cite{Inchauspe2022,Heisenberg2023} in the LISANode simulator will allow more realistic tests in the future.
Furthermore, our best estimates of instrument noise are constantly changing and may require updates of some of the presented tests. 
Finally, we have not yet considered all possible GW source types. E.g., we have not yet investigated extreme mass ratio inspirals or the cosmological SGWB.

Altogether, the TTL subtraction scheme has been proven successful in all investigated test cases.
However, due to ongoing updates on the instrument and signal modeling side, the subtraction scheme remains to be continuously monitored.

%% file: content/acknowledgements.tex
\begin{acknowledgments}
\label{sec:acc}
    The authors thank G.~Müller for valuable feedback and discussions.
    We also appreciate the helpful conversations with H.~Wegener on the content of this paper and the help of O.~Hartwig with all the questions we encountered in the process of writing the paper.
	Furthermore, the authors wish to thank the TTL Task and Expert Groups for the lively discussions.
    The presented work heavily relied on simulation tools and LISA test data provided by others.
    We acknowledge J-B.~Bayle, O.~Hartwig, A.~Petiteau and M. Lilley for developing and maintaining the LISA simulator LISANode and thank M.~Staab for his enduring support.
    We are grateful to the LDC Working Group for providing the LDC data and to M.~Le Jeune and J-B.~Bayle for their help in implementing the data in our analysis.
    For providing the LISA GW Response package, we acknowledge the authors J-B.~Bayle, Q.~Baghi, A.~Renzini and M.~Le Jeune.
    We thank J-B.~Bayle, A.~Hees, M.~Lilley, C.~Le Poncin-Lafitte, W.~Martens and E.~Joffre for creating and maintaining the LISA Orbits simulation package and W.~Martens for providing the LISA ESA science orbits.
    Moreover, we acknowledge the LISA Simulation Expert Group for both afore mentioned LISA packages.
	Moreover, PyTDI was developed and maintained by M.~Staab and J-B.~Bayle. We wish to thank them for their efforts as well.
	Finally, the presented work would not have been possible without personal and project funding.
	The authors gratefully acknowledge support by the Deutsches Zentrum für Luft- und Raumfahrt (DLR, German Space Agency) with funding of the Federal Ministry for Economic Affairs and Climate Action (BMWK) with a decision of the Deutsche Bundestag (DLR Project Reference No.\ FKZ~50OQ2301 based on funding from FKZ~50OQ1801).
	Additionally, G.~Wanner acknowledges funding by Deutsche Forschungsgemeinschaft (DFG) via its Cluster of Excellence QuantumFrontiers (EXC 2123, Project ID 390837967) and support from the DFG Cluster of Excellence PhoenixD (EXC 2122, Project ID 390833453). She thanks both clusters for the opportunities and excellent scientific exchange via the Topical Group ``Optical Simulations" and the Research Area~``S" on Simulations.
    M.~Hartig acknowledges the support by the PRIME program of the German Academic Exchange Service (DAAD) with funds from the German Federal Ministry of Education and Research (BMBF).
	For the figures in the manuscript we use a color palette by P.~Tol. We hope that the colors will be distinguishable by all readers. 
\end{acknowledgments}

%% file: content/appendix.tex
\section{Jitters and noises used in our analysis}
\label{app:noise_and_jitter}

In this section, we provide the jitter and noise ASDs considered in our analysis. 
The jitters of the SC and the MOSAs correspond to the ones introduced in \cite{LISANode,Paczkowski2022}.
\begin{eqnarray}
  S^{1/2}_{\mathrm{SC},\alpha} &=& 5\,\mathrm{nrad}/\sqrt{\mathrm{Hz}}\
  \sqrt{ 1+ \left(\frac{0.8\,\mathrm{mHz}}{f}\right)^4 } \label{eq:SC_jitter} \\
  S^{1/2}_{\mathrm{MOSA},\varphi} &=& 2\,\mathrm{nrad}/\sqrt{\mathrm{Hz}}\
  \sqrt{ 1+ \left(\frac{0.8\,\mathrm{mHz}}{f}\right)^4 } \label{eq:MOSA_phi_jitter}\\
  S^{1/2}_{\mathrm{MOSA},\eta} &=& 1\,\mathrm{nrad}/\sqrt{\mathrm{Hz}}\
  \sqrt{ 1+ \left(\frac{0.8\,\mathrm{mHz}}{f}\right)^4 } \label{eq:MOSA_eta_jitter}
\end{eqnarray}
with $\alpha\in\{\varphi,\eta,\theta\}$.

The combined angular jitter of the SC and the MOSAs will be measured utilizing DWS. The DWS readout noise is given by \cite{LISANode}:
\begin{eqnarray}
  S^{1/2}_{\rm DWS} = \frac{70}{335}\,\mathrm{nrad}/\sqrt{\mathrm{Hz}}\
\end{eqnarray}

Furthermore, we included the following interferometer noises that couple into the lenght readout in the LISANode simulations \cite{LISANode,Bayle2023}.

Laser noise given in units of frequency:
\begin{eqnarray}
  S^{1/2}_{\rm laser} 
  = 30\,\mathrm{Hz}/\sqrt{\mathrm{Hz}}\
  \sqrt{1+ \left(\frac{2\,\mathrm{mHz}}{f}\right)^4}
\end{eqnarray}

TM acceleration noise (converted to displacement noise):
\begin{eqnarray}
  S^{1/2}_{\rm acc} 
  &=& \frac{2.4}{(2\pi\,f)^2}\,\mathrm{fm}/\mathrm{s}^2/\sqrt{\mathrm{Hz}}\ \nonumber\\
  && \cdot\ \sqrt{1+ \left(\frac{0.4\,\mathrm{mHz}}{f}\right)^2}\ 
  \sqrt{1+ \left(\frac{8\,\mathrm{mHz}}{f}\right)^4}\
\end{eqnarray}

Readout noises in the different interferometers:
\begin{eqnarray}
  S^{1/2}_{\rm ro} 
  = A_{\rm ro}\
  \sqrt{ 1+ \left(\frac{2\,\mathrm{mHz}}{f}\right)^4 }
\end{eqnarray}
with $A_{\rm ro}= \{6.35\,\mathrm{pm}/\sqrt{\mathrm{Hz}},\, 1.42\,\mathrm{pm}/\sqrt{\mathrm{Hz}},\, 3.32\,\mathrm{pm}/\sqrt{\mathrm{Hz}}\,\}$ for the inter-satellite interferometer, the TM interferometer and the reference interferometer respectively.

Telescope path length noise: 
\begin{eqnarray}
  S^{1/2}_{\rm tel} = \frac{2\pi\,f}{c} A_{\rm tel}
\end{eqnarray}
with $A_{\rm tel}= \{1\,\mathrm{fm}/\sqrt{\mathrm{Hz}},\, 2\,\mathrm{fm}/\sqrt{\mathrm{Hz}},\, 1.5\,\mathrm{fm}/\sqrt{\mathrm{Hz}}\,\}$ for the incoming light, the outgoing light and the correlated part respectively.
Note that these numbers are smaller than current estimates, compare e.g.~\cite{Wegener2024}.

Fiber (or back-link) noise:
\begin{eqnarray}
  S^{1/2}_{\rm fib} 
  = 3\,\mathrm{pm}/\sqrt{\mathrm{Hz}}\
  \sqrt{ 1+ \left(\frac{2\,\mathrm{mHz}}{f}\right)^4 }
\end{eqnarray}

\section{TTL coupling estimation for a reduced fit frequency range}
\label{app:GW_200mHz}

\begin{figure}
\includegraphics[width=\columnwidth]{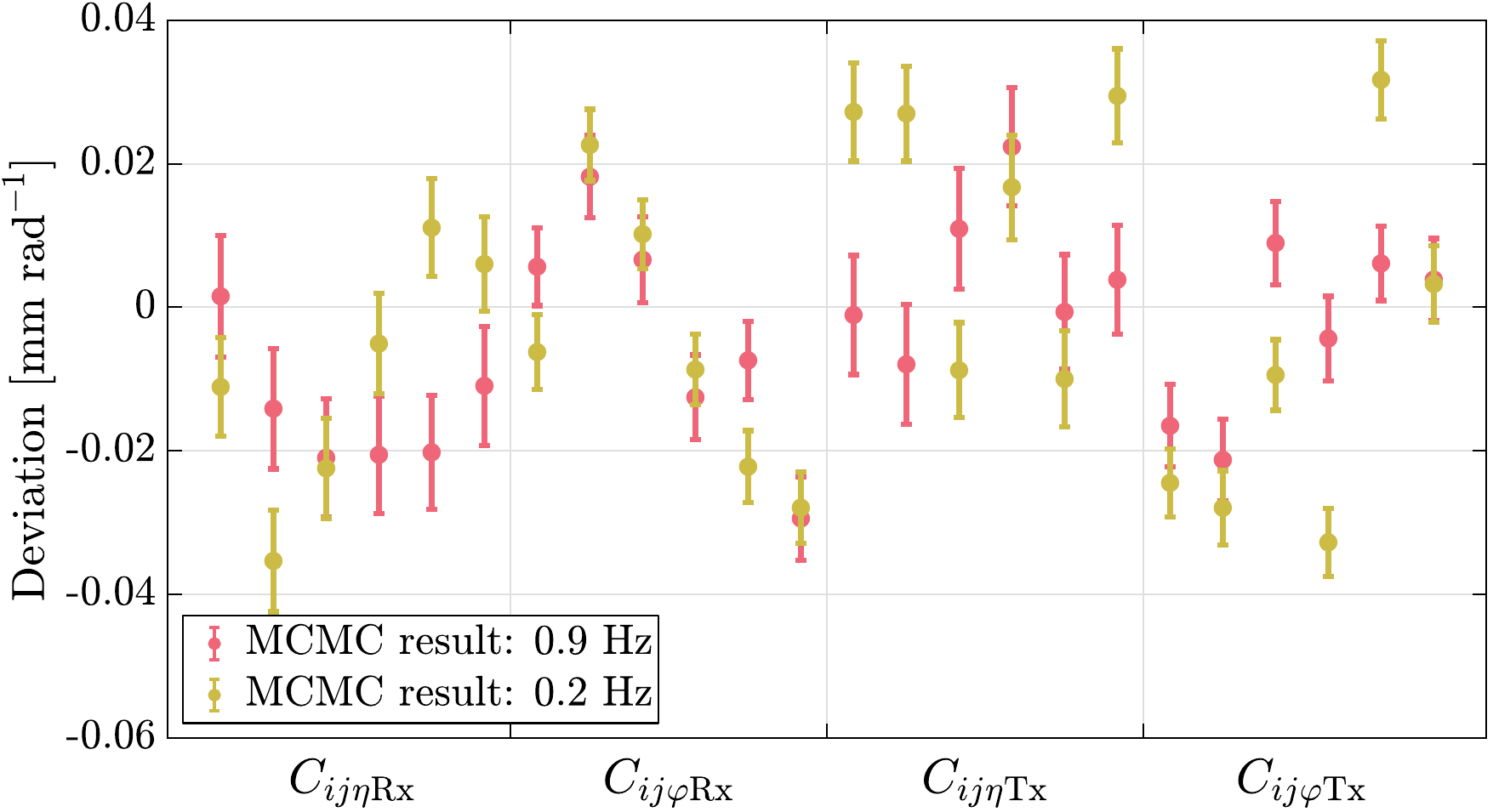} 
  \caption{Deviations of the estimated coupling coefficients from the true values in the case of the merger signal MBHB2.
  Light-red: Fit range up to 0.9\,Hz (compare figure~\ref{fig:coeff_sangria_MBHB}).
  Yellow: Fit range up to 0.2\,Hz.
  The order of the coefficient indices is $ij\in\{12,13,23,21,31,32\}$.
  }
  \label{fig:coeff_sangria_MBHB_200mHz}
\end{figure}

Throughout the paper, we have considered a fit frequency range from 3\,mHz to 0.9\,Hz. 
Here, we demonstrate that the MCMC coefficient estimation yields comparable results when the upper limit of the fit range is restricted to 0.2\,Hz.
We show this for the example of the merger MBHB2 introduced in section~\ref{sec:GW_MBHB}.
For the day considered, the ASD of the MBHB signal is above the TTL coupling noise below 12\,mHz.
The deviations of the estimated coupling coefficients for both fit frequency ranges are below the 0.1\,mm/rad requirement, see figure~\ref{fig:coeff_sangria_MBHB_200mHz}.